\documentclass[lettersize,journal]{IEEEtran}
\usepackage{amsmath,amsfonts}
\usepackage{algorithmic}
\usepackage{algorithm}
\usepackage{array}
\usepackage{textcomp}
\usepackage{stfloats}
\usepackage{url}
\usepackage{verbatim}
\usepackage{graphicx}
\usepackage{cite}
\usepackage{amsthm}
\usepackage{booktabs,tabularx}
\usepackage[normalem]{ulem}
\usepackage{xcolor}
\usepackage{multirow}
\usepackage{subfigure}

\newtheorem{problem}{Problem}
\newtheorem{theorem}{Theorem}
\newtheorem{corollary}{Corollary}
\newtheorem{definition}{Definition}
\newtheorem{example}{Example}
\newtheorem{lemma}{Lemma}
\newtheorem{proposition}{Proposition}
\newtheorem{observation}{Observation}

\begin{document}

\title{Certifying the Error-Detection Value of Functional Dependencies without Ground Truth}

\author{Xiaolong~Wan, Xixian~Han
\thanks{The authors are with the School of Computer Science and Technology, Harbin Institute of Technology, China. (e-mail: wxl@hit.edu.cn, hanxx@hit.edu.cn)}
}

\maketitle

\begin{abstract}
	Functional dependencies are used to flag erroneous cells, but without error labels the value of these flags is unknown. Dependency strength cannot settle it, because identical data and flags can have positive value in one clean state and negative value in another. We treat this value as partially identified and certify a fixed detector when its value is positive in every clean state compatible with the observation under an error model and budget. The largest such budget is the certification radius. For a detector that flags departures from the unique mode of each class, the worst case reduces to a 0/1 knapsack over class flips, yielding sharp bounds on true positives and the exact radius in polynomial time. Allowing bounded legitimate exceptions in the clean data, we derive an exact threshold for the existence of a positive certificate and compute the radius with the same knapsack. Across five relations and 21,981 candidate dependencies, rules with nearly equal $g_3$ and flag rate can have certification margins differing tenfold. Error injection confirms sound certification and shows that, at equal error counts, changing the organization of errors reverses certification outcomes for every tested rule.
\end{abstract}

\section{Introduction}\label{sec:intro}

\IEEEPARstart{F}{unctional} dependencies are a standard tool for detecting errors in relational data~\cite{DBLP:conf/icde/BohannonFGJK07,DBLP:conf/icde/ChuIP13,DBLP:books/acm/IlyasC19}. When a dependency $X \to A$ is expected to hold but does not, tuples that agree on $X$ may disagree on $A$, and a detector uses these disagreements to flag potentially erroneous cells. Acting on the flags has a cost. Each flagged cell must be inspected or repaired, and a flag on a correct value is a false alarm. How much the flags are worth, which we call their value, therefore depends on how many of them mark actual errors, and on real data this number is unknown. Counting the errors requires labeling every cell as correct or erroneous against external sources or by manual inspection, and even benchmarks with semantically annotated dependencies~\cite{parciak_2023_8098909} lack such labels. Without labels, rules are judged by quantities computed from the observed relation, most commonly dependency strength.

Measures such as $g_3$~\cite{DBLP:journals/tcs/KivinenM95} quantify how far the observed relation is from satisfying the rule, and comparative studies~\cite{DBLP:journals/is/GiannellaR04,DBLP:journals/vldb/ParciakWHNPV25} examine many alternatives. Strength describes how well a rule fits the data, not the value of its flags. Whether it nevertheless tracks value can be tested only on data with known errors, and on injected errors it does not reliably predict the value of a rule as a detector~\cite{wan2026strengthvaluerethinkingapproximate}. What is missing is a statement about the value itself that can be derived from the observation without labels.

The difficulty is that the observation does not determine the value. Consider a relation of 13 tuples in which a detector for $\mathit{Zip} \to \mathit{City}$ flags three City cells (Table~\ref{tab:running}). If the observation contains only the few errors needed to explain its conflicts, Alpha and Gamma are the correct cities of Z1 and Z2. Every flag then marks an error, and the value of the flags, i.e., the loss they avoid compared with flagging nothing, is positive. A few more errors admit a different clean state in which Delta is the correct city of Z2. The two flagged Delta cells are then correct and the unflagged Gamma cells are wrong, so only the flag on Beta still marks an error. When a missed error and a false alarm cost the same, the value is now negative, although the data and the flags are unchanged.

\begin{table}[t]
	\centering
	\caption{A relation with $\mathit{Zip} \to \mathit{City}$. Tuples whose City
		cell is flagged are marked with an asterisk.}
	\label{tab:running}
	\begin{tabular}{llll}
		\toprule
		Zip & City & Tuples & Count \\
		\midrule
		Z1 & Alpha   & $t_1, t_2, t_3, t_4$         & 4 \\
		Z1 & Beta    & $t_5^{*}$                    & 1 \\
		\midrule
		Z2 & Gamma   & $t_6, t_7, t_8, t_9$         & 4 \\
		Z2 & Delta   & $t_{10}^{*}, t_{11}^{*}$     & 2 \\
		\midrule
		Z3 & Epsilon & $t_{12}$                     & 1 \\
		Z3 & Zeta    & $t_{13}$                     & 1 \\
		\bottomrule
	\end{tabular}
\end{table}

The value is thus partially identified~\cite{manski2003partial,molinari2020microeconometrics}, known only as a member of the set of values over all clean states compatible with the observation. We fix an error model, which states what the clean data look like and how they may be corrupted, together with a budget $k$ on the number of corrupted cells. The flags are certified at budget $k$ when their value is positive in every compatible clean state. A larger budget admits more states, and the certification radius $k_\star$ is the largest budget at which the certificate still holds. We write $k_0$ for the fewest errors with which the error model explains the observation, four in the example above, one each in Z1 and Z3 and two in Z2. We call $k_\star - k_0$ the margin, the number of errors that the certificate tolerates beyond this minimum. If the clean data satisfy the error model and contain at most $k_\star$ errors, the flags have positive value, and no label enters this conclusion.

The example uses the mode-based detector, which flags every value of $A$ that differs from the unique most frequent value among the tuples sharing its $X$-value, and nothing in a group without a unique most frequent value. When the clean data satisfy the dependency and only right-hand-side cells are corrupted, the worst case over the compatible clean states arises from reinterpreting groups, as taking Delta instead of Gamma for Z2 does in the example. A reinterpretation costs additional errors and turns some flags into false alarms, and within a group only its second most frequent value needs to be considered. Choosing which groups to reinterpret within the budget is a 0/1 knapsack problem, whose solution gives a sharp lower bound on the number of correct flags at every budget and the exact certification radius in polynomial time. Groups whose two most frequent values are nearly tied offer inexpensive reinterpretations that invalidate many flags.

Clean data may also violate a dependency legitimately, as when a ZIP code spans two cities. We therefore also admit clean states with at most $\ell$ legitimate exceptions, which can be made to satisfy the dependency by changing at most $\ell$ cells of $A$. Some feasible budget certifies a positive value exactly when $\ell$ stays below a threshold determined by the number of flags and the relative costs of missed errors and false alarms. Below this threshold the same knapsack yields the certification radius, with the threshold lowered by $\ell$.

On five benchmark relations we evaluate 21,981 candidate dependencies. Nearly half of them are certified at every feasible budget. For the others, the margin usually lies near half the number of flags. Yet 1,203 candidates have a margin of zero, and many pairs whose $g_3$ and flag rate differ by less than 5\% have margins that differ tenfold or more, so $g_3$ and the flag rate alone do not determine the margin.

Injecting errors into clean data makes the true value known, and no certified run has a non-positive true value, as the theory guarantees. At equal error counts, changing how errors are organized reverses the certification outcome on every tested rule. Each run whose errors repeat one wrong value within a group is paired with a random run that corrupts exactly as many cells. On 15 of the 17 tested rules, every random run is certified, whereas every paired run in which the repeated value covers 70\% of each group it enters is a detection failure, in which the flags have no positive value. On the two remaining rules, whose left-hand sides are near keys, most random errors fall into groups of a single tuple, where they consume budget without raising flags. Every random run is then an identification failure, in which the flags have a positive value that the certificate cannot establish, whereas every paired run at 30\% coverage is certified. A certification takes well under a millisecond in the median, and the slowest instance at ten million rows takes 16.3 seconds.

Our contributions are as follows.
\begin{itemize}
	\item We formulate label-free certification of the value of a fixed dependency-based detector as a partial identification problem, with the certification radius as the largest certified error budget (Section~\ref{sec:problem}).
	\item For the mode-based detector we give a sharp worst-case bound that yields the exact radius in polynomial time, and extend it to clean data with at most $\ell$ legitimate exceptions, where the bound stays sharp and some budget is certified exactly when $\ell$ is below a threshold (Sections~\ref{sec:mode} and~\ref{sec:limits}).
	\item On observed and injected data we show that certificates carry information beyond $g_3$ and flag rate, and that at equal error counts the organization of the errors reverses the certification outcome on every tested rule (Section~\ref{sec:exp}).
\end{itemize}
Section~\ref{sec:related} discusses related work, and Section~\ref{sec:conclusion} concludes.

\section{Problem Formulation}\label{sec:problem}

This section states the certification problem in general form, before any detector or error model is chosen.

\subsection{Dependency-Based Error Detection}\label{sec:problem:detection}

We consider a relation $r$ over a schema $R$ and one functional dependency $\varphi : X \to A$ at a time, with $X \subseteq R$ and a single attribute $A \in R \setminus X$. Each tuple has a unique identifier, which distinguishes tuples even when their values are identical.

Following recent work on approximate dependencies~\cite{DBLP:journals/vldb/ParciakWHNPV25}, we evaluate $\varphi$ only on tuples with non-null values on $X \cup \{A\}$, so that null semantics do not add a second source of uncertainty. These tuples form the evaluation scope $r_\varphi$ of $\varphi$, and we write $N_\varphi = |r_\varphi|$.

A cell is a pair $(t, B)$ of a tuple identifier $t$ and an attribute $B$, and $t[B]$ is its observed value. The cells $(t, B)$ with $t \in r_\varphi$ and $B \in X \cup \{A\}$ form the detection universe $\mathcal{U}_\varphi$, the cells that a detector for $\varphi$ may flag.

\begin{definition}\label{def:detector}
	A detector for $\varphi$ is a deterministic function $D_\varphi$ that maps the evaluation scope $r_\varphi$ to a set of flagged cells $F_\varphi = D_\varphi(r_\varphi) \subseteq \mathcal{U}_\varphi$. We write $f = |F_\varphi|$.
\end{definition}

The detector is fixed before its value is examined, and we assess this detector rather than search among possible detectors for $\varphi$. The definition also allows flags on left-hand-side cells, although the mode-based detector studied in the rest of the paper flags only right-hand-side cells. When $f = 0$, we say that $D_\varphi$ has \emph{no detection opportunity} on $r$, and we exclude this degenerate case from the certification problem.

\subsection{Latent Clean Worlds}\label{sec:problem:worlds}

The flags are observed, but whether a flagged cell actually holds a wrong value is not. We treat $r_\varphi$ as the result of corrupting an unobserved clean relation in cell values only, with no tuple inserted or deleted, so tuple identity survives and errors can be located by position without matching tuples.

\begin{definition}\label{def:world}
	A latent clean world for $r_\varphi$ is a relation $r^*$ over $X \cup \{A\}$ with the same tuple identifiers as $r_\varphi$, whose cells take non-null values from the attribute domains. We write $r^*(t)[B]$ for the value of the cell $(t, B)$ in $r^*$ and $r_\varphi(t)[B]$ for its observed value $t[B]$. The error set of the observation relative to $r^*$ is the set of cells on which $r_\varphi$ and $r^*$ differ,	$E(r^*) = \{\, (t, B) \in \mathcal{U}_\varphi : r_\varphi(t)[B] \neq r^*(t)[B] \,\}$.
\end{definition}

The clean value of a cell need not occur anywhere in $r_\varphi$, since corruption may overwrite every occurrence of a correct value. Section~\ref{sec:mode} shows that the worst case for the mode-based detector never requires such values.

Definition~\ref{def:world} admits $r_\varphi$ itself as a latent clean world. It has an empty error set and so fits every budget on the number of errors, and in it every flag is a false alarm, so a budget alone guarantees nothing about the flags. A guarantee requires an assumption about the structure of the clean data. If $\varphi$ is assumed to hold exactly in the clean relation, for instance, every observed violation must contain at least one erroneous cell, so the observation forces a minimum number of errors.

\begin{definition}\label{def:compatible}
	An error model for $\varphi$ is a predicate $\mathcal{A}$ on latent clean worlds. It may refer to $r_\varphi$, and it states the structural and corruption assumptions under which a latent world is considered compatible with the observation. For a budget $k \geq 0$, the set of compatible worlds is $\mathcal{R}_\varphi(k) = \{\, r^* : \mathcal{A}(r^*) \text{ holds and } |E(r^*)| \leq k \,\}$.
\end{definition}

The error models of Sections~\ref{sec:mode} and~\ref{sec:limits} keep every $X$-value fixed, so the error sets of their compatible worlds contain cells of $A$ only. The sets $\mathcal{R}_\varphi(k)$ grow with $k$. When the error model admits at least one latent clean world, two budgets delimit the range in which the choice of $k$ matters.

\begin{definition}\label{def:budget-range}
	The minimum feasible budget and the saturation budget of $\varphi$ under $\mathcal{A}$ are $k_0 = \min \{\, k : \mathcal{R}_\varphi(k) \neq \emptyset \,\}$ and $\bar{k} = \max \{\, |E(r^*)| : \mathcal{A}(r^*) \,\}$, and the feasible budget range is $\mathcal{K}_\varphi = \{\, k \in \mathbb{Z} : k_0 \leq k \leq \bar{k} \,\}$.
\end{definition}

The minimum feasible budget is the fewest errors with which the observation can be explained under $\mathcal{A}$. The saturation budget is finite because every error set lies in $\mathcal{U}_\varphi$. Raising the budget beyond $\bar{k}$ adds no world, so $\mathcal{R}_\varphi(k) = \mathcal{R}_\varphi(\bar{k})$ for all $k \geq \bar{k}$. If no latent clean world satisfies $\mathcal{A}$, the observation contradicts the error model, and that model is not considered.

\subsection{Error-Detection Value}\label{sec:problem:value}

Relative to a latent clean world $r^*$, some flags of $D_\varphi$ are correct and others are false alarms, while some errors receive no flag. The corresponding counts are $\mathrm{TP}(r^*) = |F_\varphi \cap E(r^*)|$, $\mathrm{FP}(r^*) = |F_\varphi \setminus E(r^*)| = f - \mathrm{TP}(r^*)$ and $\mathrm{FN}(r^*) = |E(r^*) \setminus F_\varphi| = |E(r^*)| - \mathrm{TP}(r^*)$.

We measure the worth of the flags by the loss they avoid. A missed error costs $c_{\mathrm{FN}} > 0$ and a false alarm costs $c_{\mathrm{FP}} > 0$, while correctly flagged errors and correctly unflagged cells incur neither cost. The baseline of flagging nothing misses every error, at a loss of $L_0(r^*) = c_{\mathrm{FN}}\,|E(r^*)|$. Acting on $F_\varphi$ instead incurs $L_D(r^*) = c_{\mathrm{FN}}\,\mathrm{FN}(r^*) + c_{\mathrm{FP}}\,\mathrm{FP}(r^*)$.

\begin{definition}\label{def:value}
	The error-detection value of $\varphi$ under $D_\varphi$ in a latent clean world $r^*$ is $V(r^*) = L_0(r^*) - L_D(r^*)$.
\end{definition}

Although the value also depends on the dependency, the detector and the costs, we write $V(r^*)$ because only the world varies in what follows.

Substituting the counts gives
\[
V(r^*) = (c_{\mathrm{FN}} + c_{\mathrm{FP}})\,\mathrm{TP}(r^*) - c_{\mathrm{FP}}\, f .
\]
The total number of errors cancels. The value therefore depends on the world only through $\mathrm{TP}(r^*)$, and because $f$ is observed it is an increasing affine function of $\mathrm{TP}(r^*)$.

\begin{observation}[Precision threshold]\label{obs:precision}
	Let $f > 0$ and $p_c = c_{\mathrm{FP}} / (c_{\mathrm{FN}} + c_{\mathrm{FP}})$. For every latent clean world $r^*$, $V(r^*) > 0$ if and only if $\mathrm{TP}(r^*)/f > p_c$.
\end{observation}

The technical problem is thus to determine how small $\mathrm{TP}(r^*)$ can be over the compatible worlds. The cost model only fixes the threshold that this lower bound must exceed.

\subsection{Identification and Certification}\label{sec:problem:certification}

The value that matters is the one in the actual clean relation, which we denote by $r^\circ$ to distinguish it from a generic latent clean world $r^*$. It is unobserved, and the error model is our only handle on it. If $\mathcal{A}(r^\circ)$ holds and $|E(r^\circ)| \leq k$, then $r^\circ \in \mathcal{R}_\varphi(k)$, but the data cannot tell which compatible world $r^\circ$ is, so $V(r^\circ)$ is known only as a member of a set.

\begin{definition}\label{def:identified}
	For $k \in \mathcal{K}_\varphi$, the identified set of the value is $\mathcal{I}_V(k) = \{\, V(r^*) : r^* \in \mathcal{R}_\varphi(k) \,\}$.
\end{definition}

This is the identified set in the sense of partial identification~\cite{manski2003partial,molinari2020microeconometrics}. It need not be an interval. We work with its endpoints, which are determined by the extremes of the true-positive count,
\begin{align*}
	\mathrm{TP}_{\min}(k) &= \min_{r^* \in \mathcal{R}_\varphi(k)} \mathrm{TP}(r^*), \\
	\mathrm{TP}_{\max}(k) &= \max_{r^* \in \mathcal{R}_\varphi(k)} \mathrm{TP}(r^*).
\end{align*}
Both extremes exist even when infinitely many worlds are compatible, because $\mathrm{TP}(r^*)$ takes values in $\{0, \ldots, f\}$, and each is attained by some compatible world. Since the value is an increasing affine function of $\mathrm{TP}(r^*)$, the sharp lower and upper bounds of $\mathcal{I}_V(k)$ are
\begin{align*}
	V_{\min}(k) &= (c_{\mathrm{FN}} + c_{\mathrm{FP}})\,\mathrm{TP}_{\min}(k) - c_{\mathrm{FP}}\, f, \\
	V_{\max}(k) &= (c_{\mathrm{FN}} + c_{\mathrm{FP}})\,\mathrm{TP}_{\max}(k) - c_{\mathrm{FP}}\, f.
\end{align*}
Identifying the value thus reduces to identifying the true-positive count. These extremes are not immediate to compute, which leads to two problems that Section~\ref{sec:mode} solves for the mode-based detector.

\begin{problem}[Identification]\label{prob:identification}
	Given $r_\varphi$, $D_\varphi$, $\mathcal{A}$ and $k \in \mathcal{K}_\varphi$,
	compute $\mathrm{TP}_{\min}(k)$ and $\mathrm{TP}_{\max}(k)$.
\end{problem}

\begin{problem}[Certification]\label{prob:certification}
	Given in addition $c_{\mathrm{FN}}$ and $c_{\mathrm{FP}}$, decide whether $V_{\min}(k) > 0$, or equivalently, by Observation~\ref{obs:precision}, whether $\mathrm{TP}_{\min}(k) > p_c f$.
\end{problem}

If $\mathcal{A}(r^\circ)$ holds and $|E(r^\circ)| \leq k$, then $r^\circ$ is compatible, so $V_{\min}(k) > 0$ certifies $V(r^\circ) > 0$ without any label. This is the sense in which we certify a dependency.

A negative answer, $V_{\min}(k) \leq 0$, has two possible causes that the data cannot separate. If $V(r^\circ) \leq 0$, we speak of a \emph{detection failure}. If $V(r^\circ) > 0$ but a compatible world with non-positive value cannot be excluded, we speak of an \emph{identification failure}. A failure to certify is therefore not a certificate of failure, which would require $V_{\max}(k) \leq 0$, so that no compatible world has positive value. Our focus is on $V_{\min}(k)$, because the decision of whether to act on a detector requires a guaranteed benefit.

Each dependency is certified separately. The worst-case worlds of different dependencies need not fit together into one clean relation, so their lower bounds need not be attained at once. Constraining the dependencies jointly could tighten the bounds, but it would turn certification into a global inference problem over all of them.

\subsection{Certification Robustness}\label{sec:problem:robustness}

Problem~\ref{prob:certification} takes the budget as given, yet an exact bound on the number of errors is rarely available. We therefore ask the converse question of how large the budget may grow before certification is lost. Because a larger budget only adds compatible worlds, $V_{\min}(k)$ is non-increasing in $k$, so once certification fails at some budget, no larger budget restores it.

\begin{definition}\label{def:radius}
	The certification radius of $\varphi$ under $D_\varphi$ and $\mathcal{A}$ is $k_\star = \max \{\, k \in \mathcal{K}_\varphi : V_{\min}(k) > 0 \,\}$,	and $k_\star$ is undefined when this set is empty.
\end{definition}

We use the term radius because the budget bounds the cell-wise Hamming distance between the observation and a latent clean world. By monotonicity, the certified budgets form an initial segment of $\mathcal{K}_\varphi$, and each dependency falls into exactly one of three cases. It is \emph{not certifiable} when $V_{\min}(k_0) \leq 0$. Then $k_\star$ is undefined, since even the smallest compatible set contains a world with non-positive value. It has a \emph{finite certification radius} when $k_0 \leq k_\star < \bar{k}$, and its value is certified up to $k_\star$ and not beyond. It is \emph{certified throughout the feasible range} when $k_\star = \bar{k}$, in which case the value is positive in every world the error model admits.

The radius restates the certificate without a fixed budget. If $\mathcal{A}(r^\circ)$ holds and $|E(r^\circ)| \leq k_\star$, then $V(r^\circ) > 0$, so the user need not know the number of errors but must justify this upper bound. When $k_\star$ exists, we call $k_\star - k_0$ the \emph{margin}, the corruption that the certificate tolerates beyond the minimum needed to explain the observation. The budget $k_0$ is determined by the observation and the error model, whereas $k_\star$ also depends on the detector and the costs. The radius evaluates the flags of a dependency taken as given, and the validity of the dependency in the clean data is part of the error model.

\section{Mode-Based Detection under Bounded Right-Hand-Side Corruption} \label{sec:mode}

When the clean data satisfy the dependency and only right-hand-side values are corrupted, a compatible world is fixed by one clean value of $A$ per class, and a direct search for $\mathrm{TP}_{\min}(k)$ grows exponentially with the number of classes. Since errors and true positives add up over the classes and only the mode and a runner-up value matter in each class, the worst case is a 0/1 knapsack over the classes that may switch to their runner-up.

\subsection{The Mode-Based Detector}\label{sec:mode:detector}

The detector treats the tuples that agree on $X$ group by group and depends on a group only through how often each value of $A$ occurs in it. Tuples of $r_\varphi$ that agree on $X$ form an $X$-class, and we write $C_1, \ldots, C_M$ for the classes and $n_i = |C_i|$. For a value $v$ of $A$, let $c_i(v)$ be the number of tuples in $C_i$ with $A$-value $v$. Ordering the values that occur in $C_i$ by decreasing $c_i(v)$, we write $m_i$ and $s_i$ for the first two counts, with $s_i = 0$ if only one value occurs and $s_i = m_i$ if two values share the largest count. Then $g_i = n_i - m_i$ counts the tuples whose $A$-value differs from a chosen most frequent value, and it is also the smallest number of cells of $A$ that must change for all tuples of $C_i$ to agree on $A$.

\begin{definition}\label{def:mode}
	Call $C_i$ \emph{decisive} if $m_i > s_i$, and let $a_i$ be its unique mode of $A$. The mode-based detector flags $F_\varphi = \{\, (t, A) : t \in C_i,\ C_i \text{ decisive},\ t[A] \neq a_i \,\}$. Classes with $m_i = s_i$ are \emph{tie classes} and receive no flag.
\end{definition}

The mode need not form an absolute majority, as with counts 4, 3 and 2. Each decisive class contributes the $g_i$ cells whose value differs from its mode, whereas a tie class contributes none, so $f = \sum_{i \,:\, m_i > s_i} g_i$. From here on, $D_\varphi$ denotes this detector.

The detector reads only these counts, so it needs neither labels nor a probabilistic model. In a decisive class, the fewest changes that make all tuples agree on $A$ set every cell to the mode and change exactly the $g_i$ flagged cells, which motivates the flags without presuming that the mode is the clean value. Whether it is clean depends on the latent world, and the analysis below allows the clean value of any class to differ from its mode. In a tie class, setting every cell to any of the most frequent values takes equally few changes, so the detector does not pick one arbitrarily and flags nothing.

\subsection{Bounded Right-Hand-Side Corruption}\label{sec:mode:model}

The budget bounds how much corruption there is, and the error model restricts where it can occur.

\begin{definition}\label{def:a1}
	The error model $\mathcal{A}_1$ holds for a latent clean world $r^*$ if
	\begin{enumerate}
		\item $r^*(t)[B] = r_\varphi(t)[B]$ for every $t \in r_\varphi$ and every
		$B \in X$, 
		\item $r^*$ satisfies $\varphi$.
	\end{enumerate}
\end{definition}

$\mathcal{A}_1$ imposes two restrictions. The first condition keeps $X$ uncorrupted, so every world that satisfies $\mathcal{A}_1$ has the observed classes $C_1, \ldots, C_M$ and only the clean values of $A$ remain latent. If $X$-values could be wrong, a tuple might belong to a different class in the clean data, and the classes themselves would be unknown. The second condition excludes legitimate exceptions, a requirement that Section~\ref{sec:limits} relaxes.

Otherwise $\mathcal{A}_1$ leaves the distribution of errors over the classes and values of $A$ unconstrained. A class may contain any number of errors, which may all carry one wrong value that can rival or even outnumber the clean value.

\subsection{Structure of Compatible Worlds}\label{sec:mode:worlds}

Under $\mathcal{A}_1$ a compatible world amounts to a choice of one clean value of $A$ per class.

\begin{lemma}\label{lem:decomposition}
	A latent clean world $r^*$ satisfies $\mathcal{A}_1$ if and only if there are values $b_1, \ldots, b_M$ of $A$ such that $r^*$ agrees with $r_\varphi$ on $X$ and $r^*(t)[A] = b_i$ for every $t \in C_i$. In that case $|E(r^*)| = \sum_{i=1}^{M} \bigl( n_i - c_i(b_i) \bigr)$.
\end{lemma}

\begin{proof}
	By the first condition of $\mathcal{A}_1$, $r^*$ agrees with $r_\varphi$ on $X$, so its $X$-classes are $C_1, \ldots, C_M$. The second condition then holds exactly when the tuples of each class share one value of $A$ in $r^*$. A cell $(t, A)$ with $t \in C_i$ belongs to $E(r^*)$ exactly when $t[A] \neq b_i$, and $C_i$ contains $n_i - c_i(b_i)$ such tuples.
\end{proof}

We call $(b_1, \ldots, b_M)$ the clean assignment of $r^*$. The true-positive count decomposes in the same way. In a decisive class, a flagged cell $(t, A)$ is an error exactly when $t[A] \neq b_i$. If $b_i$ is the mode $a_i$, all $g_i$ flagged cells are errors, and otherwise the $c_i(b_i)$ flagged cells that carry $b_i$ become clean. A tie class carries no flag. The contribution of $C_i$ is therefore
\[
\mathrm{TP}_i(b_i) =
\begin{cases}
	g_i, & C_i \text{ decisive and } b_i = a_i, \\
	g_i - c_i(b_i), & C_i \text{ decisive and } b_i \neq a_i, \\
	0, & C_i \text{ a tie class},
\end{cases}
\]
and $\mathrm{TP}(r^*) = \sum_{i=1}^{M} \mathrm{TP}_i(b_i)$. Both the error count and the true-positive count are sums over classes, and each term depends only on the value chosen for its own class.

Each term $n_i - c_i(b_i)$ of the error count is smallest, equal to $g_i$, when $b_i$ is a most frequent value of $C_i$, and it reaches $n_i$ when $b_i$ does not occur in $C_i$. The smallest number of errors with which the observation can be explained under $\mathcal{A}_1$ is therefore $G = \sum_{i=1}^{M} g_i$.

\begin{corollary}\label{cor:range}
	The minimum feasible budget is $k_0 = G$. If, for each class, the domain of $A$ contains a value that does not occur in that class, the saturation budget is $\bar{k} = N_\varphi$.
\end{corollary}

The sum $G$ runs over all classes, whereas $f$ runs over the decisive ones only, so $G - f = \sum_{i \,:\, m_i = s_i} g_i$ counts the errors forced in tie classes, which raise no flag. For $k \in \mathcal{K}_\varphi$, we call $S = k - G$ the excess budget, the part of $k$ beyond the $G$ errors that the observation forces.

\subsection{Sharp Bounds on True Positives}\label{sec:mode:bounds}

A compatible world can reduce the number of true positives only by taking a clean value other than the mode in some decisive class. By Lemma~\ref{lem:decomposition}, taking $b_i \neq a_i$ as the clean value of a decisive class $C_i$ costs $m_i - c_i(b_i)$ errors beyond $g_i$ and removes $c_i(b_i)$ true positives. When $s_i > 0$, a value that occurs $s_i$ times in $C_i$ is called a \emph{runner-up} value of the class. A \emph{flip} of $C_i$ takes a runner-up value instead of the mode as the clean value. It changes the clean assignment, not the observation or the flags. A flip costs $d_i = m_i - s_i$ additional errors and removes $s_i$ true positives, where $d_i \geq 1$ because $m_i > s_i$. The flippable classes are indexed by $I = \{\, i : m_i > s_i,\ s_i > 0 \,\}$, exactly the classes that carry flags. Changing the clean value of any other class removes no true positive.

\begin{lemma}\label{lem:dominance}
	For every $r^* \in \mathcal{R}_\varphi(k)$ there is $\hat{r} \in \mathcal{R}_\varphi(k)$ with $\mathrm{TP}(\hat{r}) \leq \mathrm{TP}(r^*)$ whose clean assignment takes the mode or a runner-up value in every class of $I$ and a most frequent value in every other class.
\end{lemma}

\begin{proof}
	Let $(b_1, \ldots, b_M)$ be the clean assignment of $r^*$. We modify it class by class without increasing the error count or the true-positive count.
	
	A class outside $I$ is a tie class or holds a single value of $A$, so it carries no flag and contributes no true positive whatever its clean value. Replacing $b_i$ by a most frequent value of the class brings its error count to the minimum $g_i$.
	
	A class of $I$ whose clean value is the mode $a_i$ needs no change. If $b_i \neq a_i$, then $c_i(b_i) \leq s_i$, since only the mode occurs more than $s_i$ times in the class. Replacing $b_i$ by a runner-up value changes the error count of the class from $n_i - c_i(b_i)$ to $n_i - s_i$ and its contribution to $\mathrm{TP}$ from $g_i - c_i(b_i)$ to $g_i - s_i$. Neither increases.
	
	By Lemma~\ref{lem:decomposition}, the modified assignment defines a world $\hat{r}$ that satisfies $\mathcal{A}_1$, and it has the form required by the lemma. Both counts are sums over classes and neither increased in any class, so $|E(\hat{r})| \leq |E(r^*)| \leq k$ and $\mathrm{TP}(\hat{r}) \leq \mathrm{TP}(r^*)$. Hence $\hat{r} \in \mathcal{R}_\varphi(k)$.
\end{proof}

\begin{theorem}[Sharp lower bound]\label{thm:main}
	Let $k \in \mathcal{K}_\varphi$ and $S = k - G$. Then $\mathrm{TP}_{\min}(k) = f - Q(S)$, where
	\[
	Q(S) = \max \Bigl\{\, \sum_{i \in J} s_i \;:\; J \subseteq I,\
	\sum_{i \in J} d_i \leq S \,\Bigr\}.
	\]
\end{theorem}

\begin{proof}
	For $J \subseteq I$, let $r_J$ be a world that flips the classes in $J$, keeps the mode in the other classes of $I$ and takes a most frequent value in every class outside $I$. By Lemma~\ref{lem:dominance}, $\mathrm{TP}_{\min}(k)$ is the smallest $\mathrm{TP}(r_J)$ among the worlds $r_J$ in $\mathcal{R}_\varphi(k)$. By Lemma~\ref{lem:decomposition} and the class contributions $\mathrm{TP}_i$, the world $r_J$ has $G + \sum_{i \in J} d_i$ errors and $f - \sum_{i \in J} s_i$ true positives. It therefore lies in $\mathcal{R}_\varphi(k)$ exactly when $\sum_{i \in J} d_i \leq S$, and minimizing $f - \sum_{i \in J} s_i$ over these sets gives $f - Q(S)$.
\end{proof}

Theorem~\ref{thm:main} reduces the worst case over all compatible worlds to a 0/1 knapsack~\cite{cormen2022introduction} with capacity $S$, in which each class of $I$ is an item of weight $d_i$ and value $s_i$, and $Q(S)$ is the largest reduction in true positives within the excess budget.

The upper bound on true positives is attained by $r_\emptyset$, the world in which every decisive class keeps its mode as the clean value. This world has the minimum possible error count $G \leq k$, so it is compatible at every feasible budget, and all $f$ flagged cells differ from their clean values. Since no world has more than $f$ true positives, $\mathrm{TP}_{\max}(k) = f$ and $V_{\max}(k) = c_{\mathrm{FN}}\, f$. When $f > 0$, some compatible world thus gives the flags a positive value, and the data can never show that the detector lacks positive value.

Certification rests on the lower bound instead. By Observation~\ref{obs:precision}, Theorem~\ref{thm:main} gives the certification test below.

\begin{corollary}\label{cor:certify}
	For $k \in \mathcal{K}_\varphi$ and $S = k - G$, $V_{\min}(k) > 0$ if and only if $Q(S) < (1 - p_c)\, f$. When $c_{\mathrm{FN}} = c_{\mathrm{FP}}$, so that $p_c = 1/2$, the condition becomes $Q(S) < f/2$.
\end{corollary}

\subsection{Computing the Certification Radius}\label{sec:mode:radius}

Applying Corollary~\ref{cor:certify} across the feasible range gives the radius. Since a larger excess budget admits more flip sets, $Q$ is non-decreasing in $S$, and
\[
k_\star = G + \max \{\, S \in \{0, \ldots, \bar{k} - G\} :
Q(S) < (1 - p_c)\, f \,\}.
\]
The values of $Q$ at the two ends of the range settle whether a radius exists and whether the dependency is certified throughout the feasible range.

At the lower end $S = 0$, and since every flip costs $d_i \geq 1$ additional errors, no flip fits. Hence $Q(0) = 0$ and $\mathrm{TP}_{\min}(k_0) = f$, so under $\mathcal{A}_1$ every dependency that raises a flag is certified at $k_0$.

\begin{corollary}\label{cor:exists}
	If $f > 0$, then $V_{\min}(k_0) = c_{\mathrm{FN}}\, f > 0$. Consequently, $k_\star$ exists and $k_\star \geq k_0$.
\end{corollary}

At the upper end, let $D = \sum_{i \in I} d_i$ be the cost of flipping every class in $I$. For $S \geq D$ all flips fit within the excess budget, so $Q(S) = \sum_{i \in I} s_i$. The world that flips every class in $I$ uses only observed values, satisfies $\mathcal{A}_1$ and has $G + D$ errors. Since $\bar{k}$ is the largest error count under $\mathcal{A}_1$, $\bar{k} - G \geq D$ regardless of the domain of $A$, and $\mathrm{TP}_{\min}(\bar{k}) = f - \sum_{i \in I} s_i$. Because $\mathrm{TP}_{\min}(k)$ is non-increasing in $k$, a dependency is certified throughout the feasible range exactly when $\sum_{i \in I} s_i < (1 - p_c)\, f$.

The values $Q(S)$ for $0 \leq S \leq D$ are computed by the standard dynamic program for the 0/1 knapsack~\cite{kellerer2004knapsack}, which processes the classes of $I$ one at a time and records, for each capacity from $0$ to $D$, the largest reduction in true positives achievable with the classes processed so far. Since $Q$ is constant beyond $D$, these values give the whole curve $\mathrm{TP}_{\min}(k) = f - Q(S)$ over the feasible range.

\begin{proposition}\label{prop:cost}
	The certification radius, together with $\mathrm{TP}_{\min}(k)$ for every $k \in \mathcal{K}_\varphi$, can be computed in $O(N_\varphi + |I|\, D)$ expected time and $O(N_\varphi + D)$ space. Since $D \leq N_\varphi - G$ and $|I| \leq N_\varphi$, the time is $O(N_\varphi^2)$.
\end{proposition}

\begin{proof}
	A single pass over $r_\varphi$ with hash tables keyed on $X$ and on $X \cup \{A\}$ counts the values of $A$ in every class in expected $O(N_\varphi)$ time, and $n_i$, $m_i$, $s_i$ and $a_i$ follow from these counts in $O(N_\varphi)$ time. The dynamic program uses a one-dimensional table of $D + 1$ entries, performing at most one update per class of $I$ at each capacity. Since $Q$ is constant beyond $D$, the whole curve $\mathrm{TP}_{\min}(k)$ over $\mathcal{K}_\varphi$, which has at most $N_\varphi + 1$ budgets, and the radius follow in $O(N_\varphi)$ further time. The hash tables take $O(N_\varphi)$ space. Each flip cost satisfies $d_i \leq m_i$, so $D \leq \sum_i m_i = N_\varphi - G$, and $|I| \leq M \leq N_\varphi$. The bound assumes constant-time access, hashing and comparison of the relevant attribute values.
\end{proof}

The general 0/1 knapsack problem is NP-hard under binary encoding~\cite{kellerer2004knapsack}. Here the weights are differences of counts and sum to $D \leq N_\varphi$, so the dynamic program, pseudo-polynomial in general, is polynomial in the size of the relation.

\subsection{Class Structure and the Certification Radius}\label{sec:mode:drivers}

By Theorem~\ref{thm:main}, the bound $f - Q(S)$ with $S = k - G$ depends only on $f$, $G$ and the pairs $(d_i, s_i)$ of the classes in $I$, and not on $N_\varphi$. Two dependencies that agree on $N_\varphi$, $G$ and $f$ can still differ in their pairs and hence in their certification radius.

\begin{example}[Equal strength, different certification radii]\label{ex:strength}
	Consider two relations, each with two classes of six tuples. In the first, both classes have counts $5$ and $1$. In the second, one class has counts $4$ and $2$, while the other holds a single value of $A$. Both relations have $N_\varphi = 12$, $G = 2$ and $f = 2$, but the first has two flips with $(d_i, s_i) = (4, 1)$, whereas the second has one with $(d_i, s_i) = (2, 2)$. When $c_{\mathrm{FN}} = c_{\mathrm{FP}}$, certification requires $Q(S) < 1$. Since every flip removes at least one true positive, this holds exactly when no flip fits, that is, for $S \leq 3$ in the first relation and $S \leq 1$ in the second. The radii are therefore $5$ and $3$.
\end{example}

Any measure determined by $G$ and $N_\varphi$ alone, such as the $g_3$ error~\cite{DBLP:journals/tcs/KivinenM95}, which equals $G/N_\varphi$, takes the same value on the two relations, and so does the flag rate $f/N_\varphi$, so neither can determine the certification radius. This complements the empirical finding that dependency strength need not reflect error-detection value~\cite{wan2026strengthvaluerethinkingapproximate}.

Two simple lower bounds on $\mathrm{TP}_{\min}(k)$ show which classes matter at a given budget. They use the efficiency $e_i = s_i / d_i$ of a flip, the true positives it removes per additional error, and $e_{\max} = \max_{i \in I} e_i$, or $0$ if $I$ is empty.

\begin{corollary}\label{cor:structural}
	For $k \in \mathcal{K}_\varphi$ and $S = k - G$, let $I_S = \{\, i \in I : d_i \leq S \,\}$. Then $\mathrm{TP}_{\min}(k) \geq f - \sum_{i \in I_S} s_i$ and $\mathrm{TP}_{\min}(k) \geq f - e_{\max}\, S$.
\end{corollary}

\begin{proof}
	By Theorem~\ref{thm:main}, $\mathrm{TP}_{\min}(k) = f - Q(S)$, so it suffices to bound $Q(S)$. A feasible flip set $J$ contains only classes with $d_i \leq S$, so $J \subseteq I_S$ and $\sum_{i \in J} s_i \leq \sum_{i \in I_S} s_i$. For the second bound, $\sum_{i \in J} s_i = \sum_{i \in J} e_i d_i \leq e_{\max} \sum_{i \in J} d_i \leq e_{\max} S$.
\end{proof}

By the first bound, a class whose flip cost exceeds the excess budget keeps all its flags as true positives in every compatible world. By the second, every excess budget with $e_{\max} S < (1 - p_c)\, f$ is certified. When $c_{\mathrm{FN}} = c_{\mathrm{FP}}$, certification can therefore fail at an excess budget below $f/2$ only if some class has $e_i > 1$, that is, a runner-up count $s_i$ larger than the lead $d_i$ of the mode. Both bounds can be loose, and the exact radius still requires the knapsack.

Classes differ widely in this efficiency. Counts $24$ and $1$ give $d_i = 23$ and $s_i = 1$, so a flip costs many errors to remove a single true positive and $e_i = 1/23$. Counts $14$ and $11$ give $e_i = 11/3$, and three additional errors suffice for a compatible world in which eleven flags are false alarms. The greatest threat arises near a tie. Counts $25$ and $25$ form a tie class that raises no flag, but one more tuple with the first value gives $25$ flags with $d_i = 1$ and $e_i = 25$, the highest efficiency that a class of $51$ tuples admits. Once $S \geq 1$, some compatible world turns all $25$ of these flags into false alarms, so certification requires $(1 - p_c) f > 25$, which for $c_{\mathrm{FN}} = c_{\mathrm{FP}}$ means more than $50$ flags in all.

\section{Limits of Certification under Legitimate Exceptions}\label{sec:limits}

Model $\mathcal{A}_1$ requires the clean data to satisfy the dependency exactly. This section admits up to $\ell$ legitimate exceptions in the clean data and shows that no budget certifies a positive value once $\ell$ reaches $(1 - p_c) f$, while for smaller $\ell$ the same knapsack still yields the sharp bound and the certification radius.

\subsection{Modeling Legitimate Exceptions}\label{sec:limits:model}

Clean data often violate a dependency for legitimate reasons. A ZIP code may span two cities or be reassigned to a new name over time, and a flag on the correct City of such a tuple is a false positive. We keep the first condition of $\mathcal{A}_1$, which fixes the classes $C_1, \ldots, C_M$, and relax only the second. The flags depend on the observation alone and stay unchanged, but they are now judged against a larger set of worlds.

\begin{definition}\label{def:a2}
	Let $\ell \geq 0$ be an integer. The error model $\mathcal{A}_2(\ell)$ holds for a latent clean world $r^*$ if
	\begin{enumerate}
		\item $r^*(t)[B] = r_\varphi(t)[B]$ for every $t \in r_\varphi$ and every $B \in X$,
		\item there is a vector $\mathbf{b} = (b_1, \ldots, b_M)$ of values of $A$ such that the exception set $L_{\mathbf{b}}(r^*) = \{\, (t, A) : t \in C_i,\ r^*(t)[A] \neq b_i \,\}$ has at most $\ell$ elements.
	\end{enumerate}
	We call such a vector $\mathbf{b}$ a witness vector of $r^*$, $b_i$ the canonical value of $C_i$ and the cells of $L_{\mathbf{b}}(r^*)$ legitimate exceptions.
\end{definition}

A world may have several witness vectors, which declare different cells to be exceptions. For $\ell = 0$ every class of $r^*$ carries a single value of $A$, so $\mathcal{A}_2(0)$ coincides with $\mathcal{A}_1$ and the canonical values form the clean assignment.

For a fixed world, the smallest exception set takes a most frequent value of each class in $r^*$ as its canonical value. With $m_i^*$ the largest number of tuples in $C_i$ that share a value of $A$ in $r^*$, this gives $\min_{\mathbf{b}} |L_{\mathbf{b}}(r^*)| = \sum_{i=1}^{M} (n_i - m_i^*) = N_\varphi\, g_3(r^*)$. The second equality holds because $g_3$ is the smallest fraction of tuples whose removal makes $\varphi$ hold, and with $X$ fixed, deleting tuples and changing $A$-values both take $n_i - m_i^*$ steps to make a class agree on $A$. Hence $r^*$ has a witness vector exactly when $g_3(r^*) \leq \ell / N_\varphi$, so $\mathcal{A}_2(\ell)$ requires the clean data to satisfy $\varphi$ up to $\ell$ exceptions, wherever they lie.

Under $\mathcal{A}_2(\ell)$ the compatible sets become $\mathcal{R}_\varphi(k, \ell)$, and every quantity defined from them takes $\ell$ as an added argument, as in $k_0(\ell)$, $\mathcal{K}_\varphi(\ell)$, $\mathrm{TP}_{\min}(k, \ell)$, $V_{\min}(k, \ell)$ and $k_\star(\ell)$. A world compatible for $k$ and $\ell$ stays compatible when either parameter grows, so $V_{\min}(k, \ell)$ is non-increasing in both arguments wherever the sets are non-empty.

The error set $E(r^*)$ contains the cells where the observation differs from $r^*$ and counts against the budget $k$. The exception set $L_{\mathbf{b}}(r^*)$ contains the cells where $r^*$ differs from the canonical value of their class and counts against $\ell$. A cell lies in both when its clean value is a legitimate exception and its observed value is wrong. Under $\mathcal{A}_1$, an observed value that differs from the clean value of its class must be an error, whereas under $\mathcal{A}_2(\ell)$ it may be a legitimate exception that uses no part of $k$. Since $\ell$ restricts the number but not the position of exceptions, any flagged cell may be read in this way, and the observation alone cannot tell an exception from an error. Section~\ref{sec:limits:bound} discusses where a bound on $\ell$ can come from.

\subsection{When Positive Certification Becomes Impossible} \label{sec:limits:impossible}

At the minimum budget $G$ under $\mathcal{A}_1$, each class takes a most frequent value as its clean value, so the $G$ errors include every flagged cell and every flag is a true positive. Under $\mathcal{A}_2(\ell)$ any of these cells can instead keep its observed value as a legitimate exception, saving one error. Whether some budget still certifies a positive value turns on the minimum feasible budget, which now depends on $\ell$.

\begin{lemma}\label{lem:k0-exceptions}
	Under $\mathcal{A}_2(\ell)$, the minimum feasible budget is $k_0(\ell) = \max\{0, G - \ell\}$.
\end{lemma}

\begin{proof}
	Let $r^*$ satisfy $\mathcal{A}_2(\ell)$ with witness vector $\mathbf{b}$. A cell $(t, A)$ with $t \in C_i$ and $r_\varphi(t)[A] \neq b_i$ either lies in $E(r^*)$ or keeps its observed value in $r^*$, and in the second case it lies in $L_{\mathbf{b}}(r^*)$. Class $C_i$ contains $n_i - c_i(b_i) \geq g_i$ such cells, hence $G \leq \sum_{i=1}^{M} \bigl( n_i - c_i(b_i) \bigr) \leq |E(r^*)| + |L_{\mathbf{b}}(r^*)| \leq |E(r^*)| + \ell$ and $|E(r^*)| \geq \max\{0, G - \ell\}$.
	
	For a world attaining this bound, make a most frequent observed value of each class canonical, keep $\min\{\ell, G\}$ of the $G$ cells that differ from them as legitimate exceptions and set every other one to the canonical value of its class. This world satisfies $\mathcal{A}_2(\ell)$ and has $\max\{0, G - \ell\}$ errors.
\end{proof}

Since every world that satisfies $\mathcal{A}_1$ also satisfies $\mathcal{A}_2(\ell)$, under the domain condition of Corollary~\ref{cor:range} the feasible range $\mathcal{K}_\varphi(\ell)$ runs from $k_0(\ell)$ to $N_\varphi$.

\begin{theorem}[Impossibility under exceptions]\label{thm:impossible}
	Suppose $f > 0$ and $\ell \geq (1 - p_c) f$. Then $V_{\min}(k, \ell) \leq 0$ for every $k \in \mathcal{K}_\varphi(\ell)$, so no feasible budget certifies a positive value and $k_\star(\ell)$ is undefined.
\end{theorem}

\begin{proof}
	We specialize the world in the proof of Lemma~\ref{lem:k0-exceptions}, in which the canonical value of each decisive class is its mode. The $G$ cells that differ from their canonical values are then the $f$ flagged cells and $G - f$ unflagged cells of tie classes. Keeping $\min\{\ell, f\}$ flagged cells and $\min\{\ell, G\} - \min\{\ell, f\}$ unflagged cells as legitimate exceptions gives a world $r^\ell$ with $k_0(\ell)$ errors, which belongs to	$\mathcal{R}_\varphi(k, \ell)$ for every $k \in \mathcal{K}_\varphi(\ell)$. A kept flagged cell is not an error, and every other flagged cell takes the mode as its clean value and is an error, so $\mathrm{TP}(r^\ell) = \max\{0, f - \ell\} \leq p_c f$, where the inequality uses $\ell \geq (1 - p_c) f$. Observation~\ref{obs:precision} then gives $V_{\min}(k, \ell) \leq V(r^\ell) \leq 0$.
\end{proof}

For $\ell \geq f$ the world $r^\ell$ keeps every flagged cell as a legitimate exception and has no true positives. Since true positives cannot be negative, $V_{\min}(k, \ell) = -c_{\mathrm{FP}} f$ at every feasible budget. For $\ell \geq G$ the feasible range starts at $k_0(\ell) = 0$, where the only compatible world is the observation itself, so $V_{\min}(0, \ell) = V_{\max}(0, \ell) = -c_{\mathrm{FP}} f$, a certificate of failure that $\mathcal{A}_1$ never admits.

The threshold $(1 - p_c) f$ depends only on the number of flags and the cost ratio. The deviating cells of tie classes raise $G$ but not the threshold, since keeping them as legitimate exceptions removes no true positive. When $c_{\mathrm{FN}} = c_{\mathrm{FP}}$ the threshold is $f/2$, so a rule with few flags tolerates few exceptions, however large the relation. Since $f \leq G = N_\varphi \, g_3(r_\varphi)$, a positive certificate requires $\ell / N_\varphi < (1 - p_c) f / N_\varphi \leq (1 - p_c)\, g_3(r_\varphi)$. In particular, the assumption $\ell = G$, under which the clean data violate $\varphi$ no more than the observation does, never yields a certificate.

\subsection{Certification under Bounded Exceptions}\label{sec:limits:bound}

Below the threshold of Theorem~\ref{thm:impossible}, the lower bound on true positives is again sharp and is given by the same function $Q$. Setting the legitimate exceptions of a world that satisfies $\mathcal{A}_2(\ell)$ to their canonical values yields a world that satisfies $\mathcal{A}_1$ and differs in at most $\ell$ cells. Conversely, an exception that keeps the observed value of a flagged cell turns that flag into a false alarm and saves one error, which can pay for flips.

\begin{theorem}[Sharp lower bound under exceptions]\label{thm:a2-sharp}
	For every $k \in \mathcal{K}_\varphi(\ell)$, $\mathrm{TP}_{\min}(k, \ell) = \max\bigl\{ 0,\ f - \ell - Q(k + \ell - G) \bigr\}$.
\end{theorem}

\begin{proof}
	Let $k' = \min\{k + \ell, \bar{k}\}$, which lies in $\mathcal{K}_\varphi$ because $k + \ell \geq G$ by Lemma~\ref{lem:k0-exceptions}. Since $\bar{k} - G \geq D$ (Section~\ref{sec:mode:radius}) and $Q$ is constant from $D$ on, $Q(k' - G) = Q(k + \ell - G)$, and Theorem~\ref{thm:main} gives $\mathrm{TP}_{\min}(k') = f - Q(k + \ell - G)$ under $\mathcal{A}_1$.
	
	For the lower bound, let $r^* \in \mathcal{R}_\varphi(k, \ell)$ with witness vector $\mathbf{b}$, and let $r'$ agree with $r^*$ except that every cell of $L_{\mathbf{b}}(r^*)$ takes the canonical value of its class. Each class of	$r'$ carries a single value of $A$, so $r'$ satisfies $\mathcal{A}_1$. The	two worlds differ in at most $\ell$ cells, hence $|E(r')| \leq \min\{k + \ell, \bar{k}\}$ and $r' \in \mathcal{R}_\varphi(k')$. Since each changed cell alters at most one true positive, $\mathrm{TP}(r^*) \geq \mathrm{TP}(r') - \ell \geq f - \ell - Q(k + \ell - G)$, and $\mathrm{TP}(r^*) \geq 0$.
	
	For attainment, let $r' \in \mathcal{R}_\varphi(k')$ satisfy	$\mathrm{TP}(r') = f - Q(k + \ell - G)$, and let $\mathbf{b}$ be its clean	assignment. We choose $j = \min\{\ell, |E(r')|\}$ cells of $E(r')$, true	positives first, and restore their observed values. In the resulting world	$r^*$ only the restored cells can differ from $\mathbf{b}$, so	$|L_{\mathbf{b}}(r^*)| \leq j \leq \ell$ and $r^*$ satisfies	$\mathcal{A}_2(\ell)$. Its error count $|E(r')| - j$ is at most $k$, because either $j = \ell$ and $|E(r')| \leq k + \ell$, or $j = |E(r')|$. Since true	positives are chosen first and $\mathrm{TP}(r') \leq |E(r')|$, restoring	removes $\min\{\ell, \mathrm{TP}(r')\}$ true positives. Hence $\mathrm{TP}(r^*) = \max\{0, f - \ell - Q(k + \ell - G)\}$.
\end{proof}

\begin{corollary}\label{cor:a2-radius}
	Let $f > 0$. Some feasible budget certifies a positive value under	$\mathcal{A}_2(\ell)$ if and only if $\ell < (1 - p_c) f$. In that case a	budget $k \in \mathcal{K}_\varphi(\ell)$ certifies a positive value if and	only if $Q\bigl( k - k_0(\ell) \bigr) < (1 - p_c) f - \ell$, and the	certification radius $k_\star(\ell)$ is the largest such budget.
\end{corollary}

\begin{proof}
	If $\ell \geq (1 - p_c) f$, no feasible budget certifies a positive value by Theorem~\ref{thm:impossible}. Otherwise $\ell < f \leq G$, so $k_0(\ell) = G - \ell$ and $k + \ell - G = k - k_0(\ell)$. Since $p_c f \geq 0$, Observation~\ref{obs:precision} and Theorem~\ref{thm:a2-sharp} turn $V_{\min}(k, \ell) > 0$ into the stated inequality on $Q$, which holds at $k = k_0(\ell)$ because $Q(0) = 0$.
\end{proof}

The feasible range extends at least $D$ beyond $k_0(\ell)$, so the rule is certified throughout it exactly when $\sum_{i \in I} s_i < (1 - p_c) f - \ell$. Since $Q$ does not depend on $\ell$, the dynamic program need not be rerun for different values of $\ell$.

Compared with $\mathcal{A}_1$, the exceptions can remove $\ell$ true positives on their own, so the flips must remove fewer than $(1 - p_c) f - \ell$, and the excess budget is measured from $k_0(\ell)$. Suppose that $\ell + 1 < (1 - p_c) f$ and that the rule is not certified throughout the feasible range under $\mathcal{A}_2(\ell)$. Raising $\ell$ by one lowers $k_0(\ell)$ by one and cannot enlarge the largest certified excess budget, so $k_\star(\ell + 1) \leq k_\star(\ell) - 1$, with a further drop whenever the lowered threshold crosses a step of $Q$. If $0 < (1 - p_c) f - \ell \leq s_i$ for a class with $d_i = 1$, such as a near-tie class, a single error beyond $k_0(\ell)$ breaks the certificate and $k_\star(\ell) = k_0(\ell)$.

Relying on a certificate under $\mathcal{A}_2(\ell)$ requires the user to justify that the true number of errors does not exceed $k_\star(\ell)$ and that the clean data have at most $\ell$ legitimate exceptions. The observation cannot justify the bound on $\ell$, since it is consistent with clean data that satisfy $\varphi$ exactly and with clean data equal to the observation, which have $G$ exceptions. Nor does knowing that $\varphi$ holds on almost all tuples justify it, since this bounds the fraction $\ell / N_\varphi$, whereas the threshold $(1 - p_c) f$ does not grow with the relation.

Domain knowledge helps only when it bounds the number of legitimate exceptions. An official table of the ZIP codes that span several cities names the classes that may contain exceptions, but such a code may occur in many tuples, so the table does not bound $\ell$. A list of known exceptions identifies the exceptional cells themselves. Removing their tuples and recomputing $f$, $G$ and $Q$ leaves $\ell$ to count the exceptions among the remaining tuples, and a complete list gives $\ell = 0$ and $\mathcal{A}_1$. An audit of a random sample yields a statistical bound on $\ell$, and a certificate built on it holds only with the confidence of that bound, outside the deterministic setting of this paper.

Where no bound on $\ell$ is available, the certificate can be reported as a function of $\ell$. The certified pairs $(k, \ell)$, with $\ell < (1 - p_c) f$ and $k_0(\ell) \leq k \leq k_\star(\ell)$, form a region between the line $k = G - \ell$ and the non-increasing staircase $k = k_\star(\ell)$. Given a justified bound $\ell$, the region shows the largest certified budget, and given a justified budget $k$, it shows how many exceptions the certificate can absorb.

\section{Experimental Evaluation}\label{sec:exp}

We report certificates on observed data and compare them with the truth on injected data, whose clean state is known.

\subsection{Experimental Setup}\label{sec:exp:setup}

\textit{Relations and dependencies.} We use five relations of the benchmark of Parciak et al.~\cite{parciak_2023_8098909,DBLP:journals/vldb/ParciakWHNPV25} (Table~\ref{tab:datasets}). Its 115 annotated dependencies each have a single attribute on the left-hand side. Parciak et al. screened candidate dependencies by their $g_3$ score and then by the meaning of the attributes, so we treat the annotated dependencies as plausible reference cases rather than as an unbiased sample of business rules. The annotation marks dependencies, not erroneous cells, so the benchmark provides no ground truth in the sense of this paper. Tuples with a null on $X \cup \{A\}$, written as an empty string or as ``?'', are left out of the evaluation scope.

\begin{table}[t]
	\centering
	\caption{Benchmark relations, with 115 annotated FDs in total.}
	\label{tab:datasets}
	\begin{tabular}{lrrr}
		\toprule
		Relation & Tuples & Attributes & Annotated FDs \\
		\midrule
		Adult            &    32,561 & 15 &  2 \\
		Claims           &    97,231 & 13 &  4 \\
		DBLP10k          &    10,000 & 34 & 77 \\
		Hospital         &   114,919 & 15 & 29 \\
		Tax (synthetic)  & 1,000,000 & 15 &  3 \\
		\bottomrule
	\end{tabular}
\end{table}

\textit{Enumerated candidates.} Because the annotated dependencies have single-attribute left-hand sides and, after the screening, a narrow range of $g_3$, we also enumerate candidates on all five relations to obtain a large and structurally varied population of rules. For every attribute $A$ we form all $X \to A$ with $A \notin X$ and $1 \leq |X| \leq 3$, 229,557 in total. We keep a candidate when its evaluation scope has at least 1,000 tuples and the mode-based detector raises at least one flag. Candidates with the same right-hand side, evaluation scope and induced partition receive the same certificate, and we keep the one with the shortest left-hand side, breaking ties lexicographically. The enumeration space, filters and deduplication rule were fixed before any certification result was examined.

\textit{Rules for injection.} Injection needs a clean reference, so it uses annotated dependencies that hold exactly on the observation, and the observed relation serves as the clean state. The rules were fixed before any injection was run. Left-hand sides that induce the same partition are treated as one. For each partition we take the right-hand side with the largest domain, so that random replacement rarely repeats a value within a class, and we skip partitions with fewer than 1,000 tuples in scope or with fewer than half of their tuples in classes of size at least two. This yields eleven representative rules, one per partition (Table~\ref{tab:rules}). Four RHS control rules keep the partitions of representative rules but use right-hand sides with smaller domains. Two near-key rules are added despite the second filter to examine partitions dominated by singleton classes.

\begin{table*}[t]
	\centering
	\caption{Rules used for injection. $|\mathrm{dom}(A)|$ is the number of distinct values of $A$ in the scope.}
	\label{tab:rules}
	\begin{tabular}{llrrrr}
		\toprule
		Group & Rule & $N_\varphi$ & Classes & Median class & $|\mathrm{dom}(A)|$ \\
		\midrule
		Representative & DBLP10k: p1booktitle $\to$ p1booktitlefull & 2,867 & 630 & 3 & 627 \\
		Representative & DBLP10k: p2booktitle $\to$ p2booktitlefull & 2,939 & 609 & 3 & 606 \\
		Representative & Claims: Airport Code $\to$ Airport Name & 89,207 & 387 & 18 & 387 \\
		Representative & Tax: Zip $\to$ City & 1,000,000 & 41,826 & 19 & 18,728 \\
		Representative & Hospital: Provider Number $\to$ Hospital Name & 114,919 & 4,598 & 25 & 4,400 \\
		Representative & Hospital: ZIP Code $\to$ City & 114,919 & 4,175 & 25 & 2,727 \\
		Representative & Hospital: Phone Number $\to$ ZIP Code & 114,919 & 4,585 & 25 & 4,175 \\
		Representative & Hospital: StateAvg $\to$ State & 114,919 & 1,501 & 61 & 55 \\
		Representative & Adult: education $\to$ education-num & 32,561 & 16 & 789.5 & 16 \\
		Representative & Tax: AreaCode $\to$ State & 1,000,000 & 273 & 2,489 & 52 \\
		Representative & Hospital: Measure Code $\to$ Measure Name & 114,919 & 28 & 4,578 & 28 \\
		\midrule
		RHS control & Hospital: Provider Number $\to$ Emergency Service & 114,919 & 4,598 & 25 & 2 \\
		RHS control & Hospital: Provider Number $\to$ Hospital Owner & 114,919 & 4,598 & 25 & 9 \\
		RHS control & Hospital: Provider Number $\to$ State & 114,919 & 4,598 & 25 & 55 \\
		RHS control & Tax: Zip $\to$ State & 1,000,000 & 41,826 & 19 & 52 \\
		\midrule
		Near key & DBLP10k: key1 $\to$ p1title & 10,000 & 8,918 & 1 & 8,897 \\
		Near key & DBLP10k: key2 $\to$ p2title & 10,000 & 8,707 & 1 & 8,686 \\
		\bottomrule
	\end{tabular}
\end{table*}

\textit{Corruption.} Corruption changes only cells of $A$ within the scope, so the classes stay fixed. A run has a target of $K$ errors, set to 0.5\%, 1\%, 2\%, 5\% or 10\% of $N_\varphi$. A replacement value is drawn uniformly from the values of $A$ in the clean scope, excluding the clean value of the cell. \emph{Random corruption} chooses $K$ distinct cells uniformly at random and draws a replacement value for each independently. \emph{Coherent corruption} models errors that repeat one wrong value inside a class. For a concentration $\rho$, a hotspot in class $C_i$ consists of $\lfloor \rho\, n_i \rfloor$ of its cells, which all take one replacement value drawn for the class. Classes are visited in a size-biased random order, and a hotspot is placed whole when it fits the remaining budget and skipped otherwise. We use $\rho = 0.3$ and $\rho = 0.7$. At $\rho = 0.3$ the clean value keeps the majority in every hotspot class. At $\rho = 0.7$ the injected value overtakes it, except in classes of size two and four, where the two values tie. Since hotspots are never truncated, a coherent run may realize fewer than $K$ errors. Each coherent run is therefore paired with a random run of the same realized size, besides the random run at the target $K$. Each configuration is run with 30 seeds fixed in advance, and results are reported against the realized error rate. We set $c_{\mathrm{FN}} = c_{\mathrm{FP}} = 1$, so $p_c = 1/2$.

\textit{Measures.} The certificate of a run, given by $G$, $f$, $k_\star$ and the margin $k_\star - k_0$, is computed from the observation alone. The evaluation, available only for injected data, uses the clean reference $r^\circ$ and records the true error count $k_{\mathrm{true}} = |E(r^\circ)|$, the true value $V(r^\circ)$ and the counts $\mathrm{TP}$, $\mathrm{FP}$ and $\mathrm{FN}$. Under $\mathcal{A}_1$ we have $k_0 = G$, so $H = k_{\mathrm{true}} - G$ is the excess budget $S$ at the true error count. Each run has one of four outcomes. A run with $f = 0$ offers no detection opportunity, and a run with $k_{\mathrm{true}} \leq k_\star$, equivalently $H \leq k_\star - k_0$, is certified. Otherwise it is an identification failure when $V(r^\circ) > 0$ and a detection failure when $V(r^\circ) \leq 0$.

\textit{Implementation.} All programs are written in Java (JDK 21) and run single-threaded on an Intel Core i9-13900 CPU (2.00 GHz, 24 cores) with 64 GB of main memory under Windows 11. Under the fixed seeds, the certificates, counts and outcomes are exact and hardware-independent. On random small instances the radius under $\mathcal{A}_1$ and under $\mathcal{A}_2(\ell)$ agrees with exhaustive enumeration of compatible worlds in every case, and every injected run is checked against the counting identities $\mathrm{TP} + \mathrm{FP} = f$ and $\mathrm{TP} + \mathrm{FN} = k_{\mathrm{true}}$ and against two consequences of the theory, the lower bound $\mathrm{TP} \geq f - Q(H)$ of Theorem~\ref{thm:main} at the true error count and a positive value in every certified run. A failed check aborts the whole batch, and every run reported below passes all checks.

\subsection{Certificates on Observed Data}\label{sec:exp:observed}

Every certificate reported here is conditional on the clean data satisfying the rule, exactly under $\mathcal{A}_1$ or up to $\ell$ legitimate exceptions under $\mathcal{A}_2(\ell)$.

\textit{Annotated dependencies.} Of the 115 annotated dependencies, 104 hold exactly, as expected from the screening, and three more, all on Hospital with Phone Number on the left, are violated only in tie classes, so none of them raises a flag. The remaining eight have a finite radius (Table~\ref{tab:annotated}). The three Hospital rules with margin 2 share one cause. Each has a class whose mode leads the runner-up by $d_i = 3$ and whose runner-up value holds all 25 flags of the rule, so $s_i = 25$ and three additional errors turn every flag into a false positive.

\begin{table}[t]
	\centering
	\caption{Annotated dependencies with a finite radius under $\mathcal{A}_1$.}
	\label{tab:annotated}
	\footnotesize
	\setlength{\tabcolsep}{3.5pt}
	\begin{tabular}{lrrrr}
		\toprule
		Rule & $N_\varphi$ & $G$ & $f$ & $k_\star - k_0$ \\
		\midrule
		\multicolumn{5}{l}{\emph{Claims}} \\
		Disposition $\to$ Status & 87,097 & 69 & 69 & 35,403 \\
		Status $\to$ Disposition & 87,097 & 11 & 9 & 10 \\
		\multicolumn{5}{l}{\emph{Hospital}} \\
		ZIP Code $\to$ County Name & 114,869 & 350 & 50 & 24 \\
		Phone Number $\to$ Provider Number & 114,919 & 325 & 25 & 2 \\
		Phone Number $\to$ Hospital Name & 114,919 & 325 & 25 & 2 \\
		Phone Number $\to$ Hospital Owner & 114,919 & 100 & 25 & 2 \\
		\multicolumn{5}{l}{\emph{DBLP10k}} \\
		p1booktitlefull $\to$ p1booktitle & 2,867 & 3 & 3 & 9 \\
		p2booktitlefull $\to$ p2booktitle & 2,939 & 5 & 5 & 6 \\
		\bottomrule
	\end{tabular}
\end{table}

\textit{The candidate population.} Table~\ref{tab:candidates} follows the 229,557 enumerated candidates through the filters. The scope filter acts only on DBLP10k, where many attributes are mostly null. Over the five relations, 21,981 candidates remain after all filters, and nearly half of them, 10,942, are certified at every feasible budget. By Section~\ref{sec:mode:radius} this happens exactly when $\sum_{i \in I} s_i < f/2$, so that flipping every class of $I$ removes fewer than $f/2$ true positives. It is typical of candidates whose minority values scatter over many distinct values, keeping every runner-up count small, and 88\% of these candidates have $g_3 \geq 0.1$. Their radius is the upper end of the feasible range, not a value set by the flip costs, so the rest of this subsection uses the 11,039 candidates with a finite radius.

\begin{table}[t]
	\centering
	\caption{Candidates after each filter. Kept splits into Finite, with a finite radius, and All $k$, certified at every feasible budget.}
	\label{tab:candidates}
	\setlength{\tabcolsep}{3.3pt}
	\begin{tabular}{lrrrrrrr}
		\toprule
		Relation & Enum. & Scope & No flag & Dupl. & Kept & Finite & All $k$ \\
		\midrule
		Adult    &   7,035 &       0 &    195 & 1,194 & 5,646 & 2,468 & 3,178 \\
		Claims   &   3,874 &       0 &  1,234 &   560 & 2,080 &   830 & 1,250 \\
		DBLP10k  & 204,578 & 160,714 & 28,210 & 8,135 & 7,519 & 4,706 & 2,813 \\
		Hospital &   7,035 &       0 &  3,685 & 1,457 & 1,893 &   788 & 1,105 \\
		Tax      &   7,035 &       0 &  1,219 &   973 & 4,843 & 2,247 & 2,596 \\
		\midrule
		Total    & 229,557 & 160,714 & 34,543 & 12,319 & 21,981 & 11,039 & 10,942 \\
		\bottomrule
	\end{tabular}
\end{table}

\begin{figure*}[t]
	\centering
	\includegraphics[width=\textwidth,trim=0 4pt 0 4pt,clip]{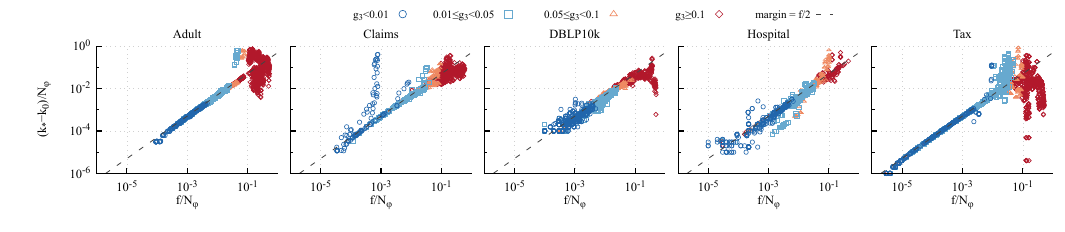}
	\caption{Normalized margin $(k_\star - k_0)/N_\varphi$ against the flag rate $f/N_\varphi$ for the candidates with a finite radius, colored by $g_3$. The dashed line marks $k_\star - k_0 = f/2$. Candidates with zero margin cannot be drawn on logarithmic axes and are omitted (Table~\ref{tab:scale}).}
	\label{fig:slack}
\end{figure*}

\textit{An empirical scale and its departures.} Figure~\ref{fig:slack} plots the normalized margin against the flag rate. Most candidates lie close to the line $k_\star - k_0 = f/2$, and among the candidates with a positive margin the correlation between the two quantities on logarithmic scales is between 0.84 and 0.97 in every relation. Only the lower side of this pattern follows from the theory, since a margin below $\lceil f/2 \rceil - 1$ requires a class with $e_i > 1$ (Section~\ref{sec:mode:drivers}). The theory gives no comparable upper bound, so the clustering near the line is an empirical finding, which makes the flag rate a reference scale for the margin. Given the flag rate, $g_3$ adds nothing consistent, since the partial correlation between $\log g_3$ and the logarithm of the margin is $-0.45$ on Claims and $-0.30$ on Tax but $-0.14$, $-0.03$ and $+0.10$ on Adult, Hospital and DBLP10k.

The scale does not replace the certificate, since a sizeable fraction of candidates departs from it by more than a factor of two (Table~\ref{tab:scale}). The most severe departures are the 1,203 candidates with zero margin, whose certificate fails at the first error beyond $k_0$, and 1,063 of them lie on DBLP10k, 22.6\% of its finite radii. Upward departures are rarer but reach a factor of ten for 4.6\% of the candidates on Tax and 2.4\% on Claims.

\begin{table}[t]
	\centering
	\caption{The ratio $(k_\star - k_0)/(f/2)$ over all candidates with a finite radius, with its median and the shares below $1/2$ and above $2$. Zero counts the candidates with zero margin.}
	\label{tab:scale}
	\begin{tabular}{lrrrrr}
		\toprule
		Relation & Finite & Median & $< 1/2$ & $> 2$ & Zero \\
		\midrule
		Adult    & 2,468 & 0.94 & 19.4\% & 15.2\% &    13 \\
		Claims   &   830 & 0.91 & 21.8\% &  9.4\% &    11 \\
		DBLP10k  & 4,706 & 0.76 & 26.9\% &  0.5\% & 1,063 \\
		Hospital &   788 & 0.85 & 18.0\% &  8.5\% &    11 \\
		Tax      & 2,247 & 0.90 & 32.8\% & 13.8\% &   105 \\
		\bottomrule
	\end{tabular}
\end{table}

\textit{Where the departures come from.} Two pairs of candidates isolate the mechanism. On Hospital, the annotated rule Phone Number $\to$ Hospital Owner and the candidate Hospital Name, County Name $\to$ Emergency Service both have $G = 100$ and $f = 25$, and their scopes differ by 50 of about 115,000 tuples, so their $g_3$ and flag rate are nearly identical. Each has a single flip with $s_i = 25 > f/2$, differing only in cost, $d_i = 3$ for the first rule and $d_i = 50$ for the second, so the margins $d_i - 1$ are 2 and 49, a factor of 24.5 apart, as in Example~\ref{ex:strength}. On DBLP10k the candidate samename, p1type $\to$ sameentity has $f = 4{,}876$ and two cheap flips, $(d_i, s_i) = (4, 1567)$ and $(3, 911)$, which together remove 2,478 true positives, more than $f/2$, for 7 additional errors, so its margin is 6. The candidate p1type, p2type $\to$ samename has nearly the same $g_3$ and flag rate, but the cheapest set of flips that removes at least $f/2$ true positives costs 249 errors, and its margin is 248. Such pairs are not rare. In every relation some candidates have a partner whose $g_3$ and flag rate differ by less than 5\% and whose margin is at least ten times larger, 550 on Adult, 358 on Tax, 21 on Hospital, 18 on Claims and 10 on DBLP10k. The flag rate thus sets a reference scale for the margin, but only the exact computation of $Q$ shows whether a given rule departs from it.

\begin{figure}[t]
	\centering
	\includegraphics[width=\columnwidth,trim=0 4pt 0 4pt,clip]{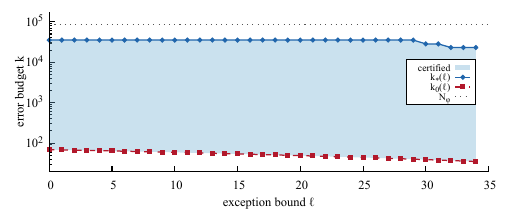}
	\caption{Certifiable region of Disposition $\to$ Status on Claims. Budgets below $k_0(\ell)$ are infeasible, budgets from $k_0(\ell)$ to $k_\star(\ell)$ are certified, and larger budgets up to $N_\varphi$ are feasible but not certified.}
	\label{fig:region}
\end{figure}

\textit{The certifiable region.} Under $\mathcal{A}_2(\ell)$ the certificate also depends on $\ell$, and Figure~\ref{fig:region} shows the certified pairs $(k, \ell)$ for Disposition $\to$ Status, the annotated rule with the largest margin. Disposition takes three values in scope, and the three classes have $(d_i, s_i) = (23{,}150, 3)$, $(28{,}426, 5)$ and $(35{,}404, 40)$, so $Q$ steps from 0 to 3, 5 and 40 at these excess budgets. By Corollary~\ref{cor:a2-radius}, the budget $k_0(\ell) + S$ is certified exactly when $Q(S) < f/2 - \ell$. For $\ell \leq 29$ this threshold exceeds 5 and only the flip of the third class breaks the certificate, so the region keeps a width of 35,404 budgets while $k_0$ and $k_\star$ both move down by one per unit of $\ell$. At $\ell = 30$ the threshold drops to 4.5 and the flip of the second class suffices, so $k_\star$ falls from 35,443 to 28,464. At $\ell = 32$ the flip of the first class suffices, and $k_\star$ falls to 23,186. From $\ell = 35$ on the threshold is negative, so not even $S = 0$ is certified. The four annotated Hospital rules show only the first pattern, a region that keeps its width and shifts down with $\ell$ until it vanishes.

\subsection{How the Organization of Errors Shapes Certification} \label{sec:exp:injection}

Table~\ref{tab:outcomes} summarizes the injected runs. The organization of the errors decides the outcome even at a fixed number of errors. In all 2,250 pairs on the fifteen rules outside the near-key group, the coherent run with $\rho = 0.7$ is a detection failure, whereas its paired random run with exactly as many errors is certified. On the two near-key rules the comparison goes the other way, with the coherent run at $\rho = 0.3$ certified and its paired random run an identification failure in all 300 pairs.

\begin{table}[t]
	\centering
	\caption{Outcomes of the injected runs over five target rates and 30 seeds per rule. Rand.\ is random corruption at the target rate, with the same outcomes as the paired random runs, and Coh.\ $\rho$ is coherent corruption. C is certified, I an identification failure and D a detection failure.}
	\label{tab:outcomes}
	\begin{tabular}{lrrrr}
		\toprule
		Group & Rules & Rand. & Coh.\ 0.3 & Coh.\ 0.7 \\
		\midrule
		Representative & 11 & 1,650 C & 1,650 C & 1,650 D \\
		RHS control    &  4 &   600 C &   600 C &   600 D \\
		Near key       &  2 &   300 I &   300 C &   300 D \\
		\bottomrule
	\end{tabular}
\end{table}

\textit{Decomposing the gap.} A certificate is conservative because the error count is unknown and because it takes the worst case at that count. The gap $\mathrm{TP} - \mathrm{TP}_{\min}(k_{\mathrm{true}})$ isolates the second cause. In class $C_i$ let $h_i$ be the number of erroneous cells minus $g_i$, which is non-negative by Lemma~\ref{lem:decomposition} since the clean reference gives each class a single clean value. Let $H = \sum_i h_i$, with $H_{\mathrm{single}}$ summing over singleton classes and $H_{\mathrm{multi}}$ over the others. Since $\mathrm{TP}_{\min}(k_{\mathrm{true}}) = f - Q(H)$ by Theorem~\ref{thm:main} and $\mathrm{TP} = f - \mathrm{FP}$, the gap decomposes as $\mathrm{TP} - \mathrm{TP}_{\min}(k_{\mathrm{true}}) = \bigl[ Q(H) - Q(H_{\mathrm{multi}}) \bigr] + \bigl[ Q(H_{\mathrm{multi}}) - \mathrm{FP} \bigr]$. Both terms are non-negative, the first since $Q$ is non-decreasing and $H_{\mathrm{multi}} \leq H$. For the second, let $J$ collect the classes of $I$ whose clean value is not the mode and occurs $c_i \geq 1$ times. A false positive is a flagged cell that holds the clean value of its class, so the false positives of $C_i$ are exactly these $c_i$ cells, and $c_i \leq s_i$. Class $C_i$ then has $h_i = m_i - c_i \geq d_i$, and since the classes of $J$ carry flags and are not singletons, $\sum_{i \in J} d_i \leq H_{\mathrm{multi}}$. Flipping the classes of $J$ thus fits within $H_{\mathrm{multi}}$ and removes at least $\mathrm{FP}$ true positives, so $\mathrm{FP} \leq Q(H_{\mathrm{multi}})$. The first term is a hidden-budget effect. Errors in singleton classes count toward $k_{\mathrm{true}}$ and raise no flag, so the worst case can spend this budget against the flags while the truth does not. The second term is an organization effect, measuring how far the errors in larger classes are from the arrangement that hurts the detector most.

\begin{figure*}[t]
	\centering
	\includegraphics[width=\textwidth,trim=0 4pt 0 4pt,clip]{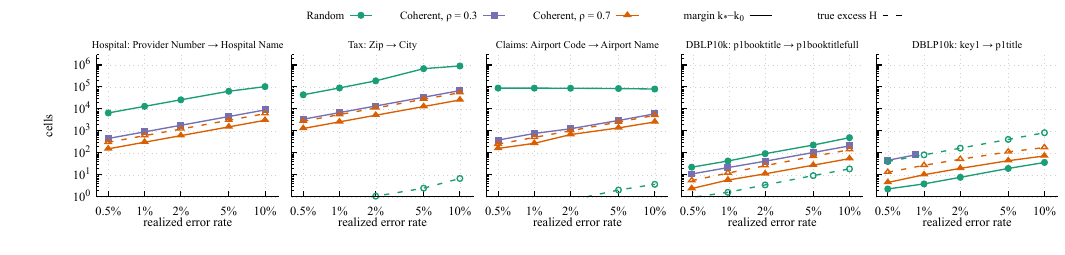}
	\caption{Mean margin $k_\star - k_0$ (solid) and mean true excess $H = k_{\mathrm{true}} - G$ (dashed) over 30 seeds, against the realized error rate, for five rules. A run is certified exactly when its excess does not exceed its margin, which is decided per run, not from the means. Zero means are omitted on the logarithmic axes.}
	\label{fig:injection}
\end{figure*}

\textit{Random corruption.} Random corruption is certified on all fifteen rules outside the near-key group, at every rate and seed, since the margin stays far above the excess $H$ (Figure~\ref{fig:injection}). Averaged over the seeds, the margins exceed $f/2$ by a factor of at least three, and by up to 400 at the lowest rate. A random error usually joins its class as an isolated minority value, which gives the class a flip with $s_i = 1$ and $d_i = n_i - 2$, costing many errors to remove a single true positive. On Claims, whose classes are large, the margin stays close to $N_\varphi$ at every rate. The gap is also small. On rules without singleton classes $H$ is zero or nearly so. Every error is then forced by the observation, and the worst case at $k_{\mathrm{true}}$ coincides with the truth. Among the rules with a gap of at least one at the 10\% rate (Table~\ref{tab:gap}), the gap stays below 19 true positives outside the near-key group, mostly from the hidden-budget term. The bound is thus nearly exact at the true error count, and what a user lacks is knowledge of that count.

\begin{table}[t]
	\centering
	\caption{Gap decomposition for random corruption at the 10\% rate, as means over 30 seeds. Gap is $\mathrm{TP} - \mathrm{TP}_{\min}(k_{\mathrm{true}})$, Hidden is $Q(H) - Q(H_{\mathrm{multi}})$ and Org.\ is $Q(H_{\mathrm{multi}}) - \mathrm{FP}$. Only rules with a gap of at least one are listed.}
	\label{tab:gap}
	\footnotesize
	\setlength{\tabcolsep}{1pt}
	\begin{tabular}{lrrrrrr}
		\toprule
		Rule & $H_{\mathrm{single}}$ & $H_{\mathrm{multi}}$ & Gap & Hidden & Org. & $k_\star - k_0$ \\
		\midrule
		p1booktitle $\to$ p1booktitlefull & 17.4 & 1.1 & 18.2 & 17.1 & 1.1 & 490.8 \\
		p2booktitle $\to$ p2booktitlefull & 17.5 & 1.3 & 17.7 & 16.4 & 1.3 & 584.0 \\
		Airport Code $\to$ Airport Name & 3.4 & 0.3 & 3.5 & 3.1 & 0.3 & 80,290.7 \\
		StateAvg $\to$ State & 8.1 & 0.4 & 6.0 & 5.6 & 0.4 & 103,436.5 \\
		Zip $\to$ City & 4.8 & 2.0 & 6.8 & 4.8 & 2.0 & 900,006.8 \\
		Zip $\to$ State & 4.7 & 2.8 & 10.3 & 5.9 & 4.4 & 900,007.5 \\
		key1 $\to$ p1title & 818.8 & 5.4 & 55.0 & 49.6 & 5.4 & 36.6 \\
		key2 $\to$ p2title & 777.4 & 6.3 & 67.5 & 61.2 & 6.3 & 38.1 \\
		\bottomrule
	\end{tabular}
\end{table}

\textit{Coherent corruption with $\rho = 0.3$.} Every run is certified and the gap is zero, because a hotspot of $a = \lfloor 0.3\, n_i \rfloor$ cells leaves the clean value in the majority, so every error is forced and $H = 0$. On twelve of the seventeen rules the margins are nevertheless one to two orders of magnitude smaller than those of the paired random runs. The factor is smaller on the two booktitle rules and on Provider Number $\to$ Emergency Service, and on the two near-key rules the coherent margins are larger. A hotspot class has $s_i = a$ and $d_i = n_i - 2a$, so its efficiency is at most $3/4$, and since classes without a hotspot have no flip, Corollary~\ref{cor:structural} guarantees a margin of at least $\lceil 2f/3 \rceil - 1$. Averaged over the seeds, the ratio of margin to $f/2$ ranges from 1.33 on AreaCode $\to$ State, close to the $4/3$ that this bound implies, to 2.42 on Measure Code $\to$ Measure Name. On the twelve rules, concentration thus moves the margin from far above the scale $f/2$ to close to it at an unchanged number of errors.

\textit{Coherent corruption with $\rho = 0.7$.} Every run is a detection failure. In each hotspot class that the injected value takes over, the detector flags the remaining clean cells, and tied hotspot classes raise no flag, so no flag marks an error and $\mathrm{TP} = 0$. The truth thus attains the worst case at $k_{\mathrm{true}}$, and the gap is zero. The observation looks the same as one in which the injected values are clean and every flag is correct, but since the truth is compatible at $k_{\mathrm{true}}$ and has negative value, no certificate can reach $k_{\mathrm{true}}$. Averaged over the seeds, the excess exceeds the margin by a factor between 1.1 and 2.9. On the zero-margin candidates of Section~\ref{sec:exp:observed}, a single error beyond $k_0$ admits a world in which the flags have no positive value.

\textit{Near keys.} On the two near-key rules every random run is an identification failure. The flags have positive value, but about four fifths of the errors fall into singleton classes, where they consume budget and raise no flag, so $H$ exceeds the margin. At the 0.5\% rate the mean $H_{\mathrm{single}}$ is about 40 of 50 errors while the margin is about 2, and at 10\% the hidden-budget term accounts for about 90\% of the gap (Table~\ref{tab:gap}). Coherent corruption leaves singleton classes untouched, since their hotspots are empty, so the same rules are certified under $\rho = 0.3$, with larger margins than under random corruption.

\subsection{Computational Cost}\label{sec:exp:cost}

\textit{Two stages.} A certificate is computed in two stages. The counting stage scans the scope once and returns $G$, $f$ and the flips, so its cost is linear in $N_\varphi$. The DP stage is skipped when $\sum_{i \in I} s_i < f/2$, which holds exactly when the rule is certified at every feasible budget. Otherwise the DP must find the smallest excess budget at which certification fails, $S_{\mathrm{fail}} = k_\star - k_0 + 1$, and it refines the dynamic program of Section~\ref{sec:mode:radius} in two ways that leave the radius unchanged. It merges flips with equal $(d_i, s_i)$ and splits each group into parts of $1, 2, 4, \ldots$ flips and a remainder, which leaves $B$ parts in total. It starts at capacity $\min\{64, D\}$, where $D$ is the cost of flipping every class of $I$, and doubles the capacity until it covers $S_{\mathrm{fail}}$, so the final capacity satisfies $C \leq \min\{D, \max\{64, 2S_{\mathrm{fail}} - 2\}\}$. The capacities of successive rounds grow geometrically, so the number of state updates is at most about $2BC$, against about $|I|\, D$ for the table of Proposition~\ref{prop:cost}.

\begin{table}[t]
	\centering
	\caption{Running time. Count and DP are the two certification stages over the injected runs, Region is the largest time for the region of an annotated rule over all $\ell$ (--- if none), and Batch processes all candidates from enumeration to certification.}
	\label{tab:cost}
	\footnotesize
	\setlength{\tabcolsep}{3.5pt}
	\begin{tabular}{lrrrrrrr}
		\toprule
		& & \multicolumn{2}{c}{Count (ms)} & \multicolumn{2}{c}{DP (ms)} & Region & Batch \\
		\cmidrule(lr){3-4}\cmidrule(lr){5-6}
		Relation & Runs & Median & Max & Median & Max & (ms) & (s) \\
		\midrule
		Adult    &   742 & 0.10 & 0.92 & 0.001 &  0.06 & --- &   4.7 \\
		Claims   &   450 & 0.19 & 2.23 & 0.007 &  0.45 &  17 &   7.2 \\
		DBLP10k  & 2,070 & 0.06 & 0.73 & 0.002 &  0.60 &   1 &   6.2 \\
		Hospital & 4,943 & 0.22 & 3.17 & 0.040 & 19.01 &  23 &  10.3 \\
		Tax      & 1,541 & 3.44 & 7.94 & 0.864 & 1,116 & --- & 278.4 \\
		\bottomrule
	\end{tabular}
\end{table}

\textit{Cost on the benchmark relations.} Over the 9,746 injected runs (Table~\ref{tab:cost}), the median certification spends 0.22\,ms on counting and 0.012\,ms on the DP, and each stage separately finishes within one millisecond in about 84\% of the runs. Counting takes at most 7.9\,ms, on Tax, the largest relation, whereas the DP ranges from under a microsecond to 1.1 seconds. All 30 runs above one second come from random corruption of Zip $\to$ State at the 5\% rate, where about 25,000 flips give $D \approx 683{,}000$ and a margin above 80\% of $D$, so the DP runs to full capacity. At 10\% the same rule is certified at every feasible budget and the DP does not run. Computing the region of an annotated rule over all $\ell$ takes at most 23\,ms, on ZIP Code $\to$ County Name. Processing all candidates takes 4.7 to 10.3 seconds on four relations and 278 seconds on Tax, where deduplication, which is not part of certification, takes 124 seconds and the DP 101 seconds.

\begin{figure}[t]
	\centering
	\includegraphics[width=\columnwidth]{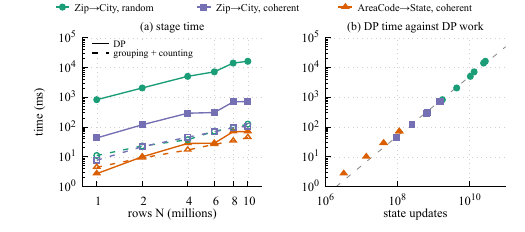}
	\caption{Running time on three injected Tax instances (5\% errors) whose classes are replicated $q$ times, so $N = q \times 10^6$. (a) Median time of each stage, where slope $1$ means linear growth. (b) DP time against state updates, with the gray line at $0.5$\,ns per update.}
	\label{fig:scale}
\end{figure}

\textit{Scaling.} To vary size while keeping the structure of the flips fixed, we replicate three injected instances of Tax at the 5\% rate with the first seed, random and coherent ($\rho = 0.3$) corruption of Zip $\to$ City and coherent corruption of AreaCode $\to$ State. Every class is copied $q$ times under new class identifiers, so $N_\varphi$, $G$, $f$, $D$ and the number of flips grow by the factor $q$ while each flip keeps its $(d_i, s_i)$. Over $q$ from 1 to 10, with medians of five repetitions (Figure~\ref{fig:scale}), grouping and counting grow by a factor of 10 to 14 and take at most 126\,ms at ten million rows, and the DP time follows the number of state updates, at 0.45 to 0.59\,ns per update on Zip $\to$ City and up to 0.84\,ns on AreaCode $\to$ State. Merging keeps $B$ small. Under random corruption of Zip $\to$ City the flips grow from 25,108 to 251,080 but fall into the same 248 distinct pairs, so the parts grow only from 1,024 to 1,803. This growth is logarithmic only because replication adds no new pairs. Here the time per update rises from 0.47 to 0.59\,ns as the DP array grows from about 5\,MB to 55\,MB, whereas the coherent instance, with an array below 5\,MB, keeps 0.45 to 0.47\,ns.

\textit{What drives the DP.} At ten million rows the three instances take 16.3, 0.82 and 0.12 seconds in total. The difference lies in $B$ and $S_{\mathrm{fail}}$. Random corruption of Zip $\to$ City gives $S_{\mathrm{fail}} = 6{,}561{,}870$, close to $D = 6{,}849{,}050$, so the DP runs to full capacity over 1,803 parts. Coherent corruption of the same rule fails at $S_{\mathrm{fail}} = 339{,}640$, which caps the capacity at 524,288 and cuts the state updates by a factor of 18. AreaCode $\to$ State fails at a similar budget with only 112 parts, which cuts the updates by a further factor of 14. Since the final capacity is less than $2S_{\mathrm{fail}}$ once $S_{\mathrm{fail}}$ exceeds 32, a small $S_{\mathrm{fail}}$ bounds the DP capacity however large the relation is, and the remaining cost depends on the number of parts. The expensive cases are rules with a large margin close to $D$, such as random corruption of Zip $\to$ City and Zip $\to$ State on Tax at the 5\% rate.

\section{Related Work}\label{sec:related}

\textit{Dependency discovery and strength.} Kivinen and Mannila~\cite{DBLP:journals/tcs/KivinenM95} measure how far a functional dependency is from holding by $g_3$, the minimum fraction of tuples whose removal makes it hold. Giannella and Robertson~\cite{DBLP:journals/is/GiannellaR04} compare such approximation measures, and Parciak et al.~\cite{DBLP:journals/vldb/ParciakWHNPV25} extend the comparison to a broad set of measures and provide the annotated benchmark~\cite{parciak_2023_8098909} used in our experiments. Caruccio et al.~\cite{DBLP:journals/tkde/CaruccioDP16} survey relaxed dependencies. On the discovery side, TANE~\cite{DBLP:journals/cj/HuhtalaKPT99} finds exact and approximate dependencies level by level over attribute partitions and accepts a dependency whose $g_3$ falls below a threshold. Bleifu{\ss} et al.~\cite{DBLP:journals/pacmmod/BleifussPBSN24} discover dependencies through hitting set enumeration, and Zhang et al.~\cite{DBLP:conf/sigmod/ZhangGR20} treat discovery in noisy data as a statistical problem. These works measure how well a rule fits the observation or search for rules that fit it well. Wan and Han~\cite{wan2026strengthvaluerethinkingapproximate} show on injected errors, whose locations are known, that such strength does not reliably predict the value of the rule as an error detector. The present paper needs no known errors and certifies that value from the observation itself, and Section~\ref{sec:exp:observed} finds that, given the flag rate, the association between $g_3$ and the certified margin varies in sign and size across relations.

\textit{Constraint-based error detection and diagnosis.} Conditional functional dependencies~\cite{DBLP:conf/icde/BohannonFGJK07} extend functional dependencies with patterns of constants and detect their violations with SQL queries, and Kaminsky et al.~\cite{DBLP:journals/pvldb/KaminskyPN24} maintain the violations of denial constraints incrementally as the data change. Holistic data cleaning~\cite{DBLP:conf/icde/ChuIP13} gathers the violations of many denial constraints into one conflict hypergraph and repairs them jointly with few changes, and NADEEF~\cite{DBLP:conf/sigmod/DallachiesaEEEIOT13} provides a common interface through which such rules are specified, checked and repaired. Data X-Ray~\cite{DBLP:conf/sigmod/WangDM15} starts from data elements known to be erroneous and explains them by features they share. Abedjan et al.~\cite{DBLP:journals/pvldb/AbedjanCDFIOPST16} evaluate error detection tools on data with ground truth, and Raha~\cite{DBLP:conf/sigmod/MahdaviAFMOS019} and HoloDetect~\cite{DBLP:conf/sigmod/HeidariMIR19} learn error detectors from limited supervision. Ilyas and Chu~\cite{DBLP:books/acm/IlyasC19} survey this area. These works develop methods for handling data errors or evaluate their empirical performance. Before any repair, and without labeled errors, we ask whether the flags of the mode-based detector have positive value in every clean state compatible with the observation within an error budget.

\textit{Repairs and consistent query answering.} Arenas et al.~\cite{DBLP:conf/pods/ArenasBC99} define a repair as an instance that satisfies the constraints and differs minimally from the database under set inclusion, and call an answer consistent when it holds in every repair. Bertossi~\cite{DBLP:series/synthesis/2011Bertossi} surveys the semantics and complexity of this approach. Pfandler and Sallinger~\cite{DBLP:conf/ijcai/PfandlerS15} bound the distance between a repair and the database by a fixed constant $b$, since inconsistencies are often limited to a small portion of a large database. For functional dependencies, where a repair deletes tuples, the combined complexity of this problem then drops from $\Pi_2^P$-complete to NP-complete. Our error budget rests on the same premise, and our certificate is likewise a statement about every world in a bounded family. The family and the statement differ. Compatible worlds change only cells of $A$, keeping tuples and classes fixed, and need not be minimal. Under $\mathcal{A}_2(\ell)$ the clean state may itself violate the rule in up to $\ell$ cells. The statement concerns the value of a detector rather than a query answer, and instead of deciding it at one bound we compute the largest budget $k_\star$ at which it holds.

\textit{Performance evaluation without ground truth.} Platanios et al.~\cite{DBLP:conf/nips/PlataniosPMH17} estimate the accuracy of several classifiers from unlabeled data by combining their agreement with logical constraints among the predicted labels. Polo et al.~\cite{DBLP:conf/nips/PoloMYBS24} evaluate models trained under weak supervision, where true labels are never observed. They treat the performance metrics as partially identified and bound accuracy, precision, recall and F1 by what the weak labels reveal about the joint distribution. The value of the flags is likewise partially identified. The observation does not determine $V(r^\circ)$, so a guarantee must hold over all states compatible with it. Here that set is fixed by the classes of the dependency, the assumption $\mathcal{A}_1$ or $\mathcal{A}_2(\ell)$ and the budget, and its worst case is computed exactly by a knapsack over class flips. The result is a sharp worst-case bound on the value together with the exact certification radius, obtained without a probabilistic model of the unobserved errors.

\section{Conclusion}\label{sec:conclusion}

This paper replaces the question of how well a functional dependency fits the observation with whether its flags provably have positive value within an error budget. For the mode-based detector the worst case over all compatible clean states reduces to a knapsack over class flips, so the certification radius is exact and computable in polynomial time. The margin $k_\star - k_0$ depends on how many errors each flip costs and how many true positives it removes rather than on the degree of violation, and legitimate exceptions lower the certification threshold one for one until no budget certifies a positive value. Rules with nearly equal $g_3$ and flag rate can have margins that differ tenfold, and at equal error counts the organization of the errors reverses the certification outcome on every tested rule. Reporting the radius and the margin next to a discovered rule therefore adds information that $g_3$ does not carry. Natural extensions are errors on the left-hand side, which make the partition itself latent, and joint constraints from several dependencies, which restrict the compatible clean states and can strengthen the guarantee.

\bibliographystyle{IEEEtran}
\bibliography{IEEEexample}

\begin{thebibliography}{10}
\providecommand{\url}[1]{#1}
\csname url@samestyle\endcsname
\providecommand{\newblock}{\relax}
\providecommand{\bibinfo}[2]{#2}
\providecommand{\BIBentrySTDinterwordspacing}{\spaceskip=0pt\relax}
\providecommand{\BIBentryALTinterwordstretchfactor}{4}
\providecommand{\BIBentryALTinterwordspacing}{\spaceskip=\fontdimen2\font plus
\BIBentryALTinterwordstretchfactor\fontdimen3\font minus
  \fontdimen4\font\relax}
\providecommand{\BIBforeignlanguage}[2]{{%
\expandafter\ifx\csname l@#1\endcsname\relax
\typeout{** WARNING: IEEEtran.bst: No hyphenation pattern has been}%
\typeout{** loaded for the language `#1'. Using the pattern for}%
\typeout{** the default language instead.}%
\else
\language=\csname l@#1\endcsname
\fi
#2}}
\providecommand{\BIBdecl}{\relax}
\BIBdecl

\bibitem{DBLP:conf/icde/BohannonFGJK07}
P.~Bohannon, W.~Fan, F.~Geerts, X.~Jia, and A.~Kementsietsidis, ``Conditional
  functional dependencies for data cleaning,'' in \emph{Proceedings of the
  {IEEE} International Conference on Data Engineering}, 2007, pp. 746--755.

\bibitem{DBLP:conf/icde/ChuIP13}
X.~Chu, I.~F. Ilyas, and P.~Papotti, ``Holistic data cleaning: Putting
  violations into context,'' in \emph{Proceedings of the {IEEE} International
  Conference on Data Engineering}, 2013, pp. 458--469.

\bibitem{DBLP:books/acm/IlyasC19}
I.~F. Ilyas and X.~Chu, \emph{Data Cleaning}, ser. {ACM} Books.\hskip 1em plus
  0.5em minus 0.4em\relax {ACM}, 2019, vol.~28.

\bibitem{parciak_2023_8098909}
\BIBentryALTinterwordspacing
M.~Parciak, S.~Weytjens, F.~Neven, N.~Hens, L.~M. Peeters, and S.~Vansummeren,
  ``Annotated benchmark of real-world data for approximate functional
  dependency discovery,'' Jun. 2023. [Online]. Available:
  \url{https://doi.org/10.5281/zenodo.8098909}
\BIBentrySTDinterwordspacing

\bibitem{DBLP:journals/tcs/KivinenM95}
J.~Kivinen and H.~Mannila, ``Approximate inference of functional dependencies
  from relations,'' \emph{Theor. Comput. Sci.}, vol. 149, no.~1, pp. 129--149,
  1995.

\bibitem{DBLP:journals/is/GiannellaR04}
C.~Giannella and E.~L. Robertson, ``On approximation measures for functional
  dependencies,'' \emph{Inf. Syst.}, vol.~29, no.~6, pp. 483--507, 2004.

\bibitem{DBLP:journals/vldb/ParciakWHNPV25}
M.~Parciak, S.~Weytjens, N.~Hens, F.~Neven, L.~M. Peeters, and S.~Vansummeren,
  ``Measuring approximate functional dependencies: a comparative study,''
  \emph{{VLDB} J.}, vol.~34, no.~4, p.~56, 2025.

\bibitem{wan2026strengthvaluerethinkingapproximate}
\BIBentryALTinterwordspacing
X.~Wan and X.~Han, ``Strength is not value: Rethinking approximate functional
  dependency measures for error detection,'' 2026. [Online]. Available:
  \url{https://arxiv.org/abs/2610.08189}
\BIBentrySTDinterwordspacing

\bibitem{manski2003partial}
C.~F. Manski, \emph{Partial Identification of Probability Distributions},
  1st~ed., ser. Springer Series in Statistics.\hskip 1em plus 0.5em minus
  0.4em\relax Springer New York, 2003.

\bibitem{molinari2020microeconometrics}
F.~Molinari, ``Microeconometrics with partial identification,'' in
  \emph{Handbook of Econometrics}.\hskip 1em plus 0.5em minus 0.4em\relax
  Elsevier, 2020, vol.~7A, pp. 355--486.

\bibitem{cormen2022introduction}
T.~H. Cormen, C.~E. Leiserson, R.~L. Rivest, and C.~Stein, \emph{Introduction
  to Algorithms}, 4th~ed.\hskip 1em plus 0.5em minus 0.4em\relax Cambridge, MA,
  USA: The MIT Press, 2022.

\bibitem{kellerer2004knapsack}
H.~Kellerer, U.~Pferschy, and D.~Pisinger, \emph{Knapsack Problems}.\hskip 1em
  plus 0.5em minus 0.4em\relax Berlin, Heidelberg: Springer, 2004.

\bibitem{DBLP:journals/tkde/CaruccioDP16}
L.~Caruccio, V.~Deufemia, and G.~Polese, ``Relaxed functional dependencies -
  {A} survey of approaches,'' \emph{{IEEE} Trans. Knowl. Data Eng.}, vol.~28,
  no.~1, pp. 147--165, 2016.

\bibitem{DBLP:journals/cj/HuhtalaKPT99}
Y.~Huhtala, J.~K{\"{a}}rkk{\"{a}}inen, P.~Porkka, and H.~Toivonen, ``{TANE:} an
  efficient algorithm for discovering functional and approximate
  dependencies,'' \emph{Comput. J.}, vol.~42, no.~2, pp. 100--111, 1999.

\bibitem{DBLP:journals/pacmmod/BleifussPBSN24}
T.~Bleifu{\ss}, T.~Papenbrock, T.~Bl{\"{a}}sius, M.~Schirneck, and F.~Naumann,
  ``Discovering functional dependencies through hitting set enumeration,''
  \emph{Proc. {ACM} Manag. Data}, vol.~2, no.~1, pp. 43:1--43:24, 2024.

\bibitem{DBLP:conf/sigmod/ZhangGR20}
Y.~Zhang, Z.~Guo, and T.~Rekatsinas, ``A statistical perspective on discovering
  functional dependencies in noisy data,'' in \emph{Proceedings of the {ACM}
  {SIGMOD} International Conference on Management of Data}, 2020, pp. 861--876.

\bibitem{DBLP:journals/pvldb/KaminskyPN24}
Y.~Kaminsky, E.~H.~M. Pena, and F.~Naumann, ``Incremental detection of denial
  constraint violations,'' \emph{Proc. {VLDB} Endow.}, vol.~18, no.~4, pp.
  1000--1012, 2024.

\bibitem{DBLP:conf/sigmod/DallachiesaEEEIOT13}
M.~Dallachiesa, A.~Ebaid, A.~Eldawy, A.~K. Elmagarmid, I.~F. Ilyas, M.~Ouzzani,
  and N.~Tang, ``{NADEEF:} a commodity data cleaning system,'' in
  \emph{Proceedings of the {ACM} {SIGMOD} International Conference on
  Management of Data}, 2013, pp. 541--552.

\bibitem{DBLP:conf/sigmod/WangDM15}
X.~Wang, X.~L. Dong, and A.~Meliou, ``Data {X-Ray}: {A} diagnostic tool for
  data errors,'' in \emph{Proceedings of the {ACM} {SIGMOD} International
  Conference on Management of Data}, 2015, pp. 1231--1245.

\bibitem{DBLP:journals/pvldb/AbedjanCDFIOPST16}
Z.~Abedjan, X.~Chu, D.~Deng, R.~C. Fernandez, I.~F. Ilyas, M.~Ouzzani,
  P.~Papotti, M.~Stonebraker, and N.~Tang, ``Detecting data errors: Where are
  we and what needs to be done?'' \emph{Proc. {VLDB} Endow.}, vol.~9, no.~12,
  pp. 993--1004, 2016.

\bibitem{DBLP:conf/sigmod/MahdaviAFMOS019}
M.~Mahdavi, Z.~Abedjan, R.~C. Fernandez, S.~Madden, M.~Ouzzani, M.~Stonebraker,
  and N.~Tang, ``Raha: {A} configuration-free error detection system,'' in
  \emph{Proceedings of the {ACM} {SIGMOD} International Conference on
  Management of Data}, 2019, pp. 865--882.

\bibitem{DBLP:conf/sigmod/HeidariMIR19}
A.~Heidari, J.~McGrath, I.~F. Ilyas, and T.~Rekatsinas, ``{HoloDetect}:
  Few-shot learning for error detection,'' in \emph{Proceedings of the {ACM}
  {SIGMOD} International Conference on Management of Data}, 2019, pp. 829--846.

\bibitem{DBLP:conf/pods/ArenasBC99}
M.~Arenas, L.~E. Bertossi, and J.~Chomicki, ``Consistent query answers in
  inconsistent databases,'' in \emph{Proceedings of the {ACM}
  {SIGACT-SIGMOD-SIGART} Symposium on Principles of Database Systems}, 1999,
  pp. 68--79.

\bibitem{DBLP:series/synthesis/2011Bertossi}
L.~E. Bertossi, \emph{Database Repairing and Consistent Query Answering}, ser.
  Synthesis Lectures on Data Management.\hskip 1em plus 0.5em minus 0.4em\relax
  Morgan {\&} Claypool Publishers, 2011.

\bibitem{DBLP:conf/ijcai/PfandlerS15}
A.~Pfandler and E.~Sallinger, ``Distance-bounded consistent query answering,''
  in \emph{Proceedings of the International Joint Conference on Artificial
  Intelligence}, 2015, pp. 2262--2269.

\bibitem{DBLP:conf/nips/PlataniosPMH17}
E.~A. Platanios, H.~Poon, T.~M. Mitchell, and E.~Horvitz, ``Estimating accuracy
  from unlabeled data: {A} probabilistic logic approach,'' in \emph{Advances in
  Neural Information Processing Systems}, 2017, pp. 4361--4370.

\bibitem{DBLP:conf/nips/PoloMYBS24}
F.~M. Polo, S.~Maity, M.~Yurochkin, M.~Banerjee, and Y.~Sun, ``Weak supervision
  performance evaluation via partial identification,'' in \emph{Advances in
  Neural Information Processing Systems}, 2024, pp. 135\,704--135\,740.

\end{thebibliography}


 


\vfill

\end{document}